\documentclass[journal]{IEEEtran}
\usepackage{graphicx}
\usepackage{makeidx}
\usepackage{fancyhdr}
\usepackage{amssymb}
\usepackage{amsmath}
\usepackage{balance}
\usepackage{array}
\usepackage{cite}
\usepackage{transparent}
\usepackage{eso-pic}
\usepackage{algorithm}
\usepackage{algpseudocode}
\usepackage{stfloats}%
\usepackage{caption}
\usepackage{subcaption}
\usepackage[utf8]{inputenc}

\ifCLASSINFOpdf
\else
\fi

\begin{document}

\title{Hybrid Beamforming for mm-Wave Massive MIMO Systems with Partially Connected RF Architecture}

\author{Mohammad~Majidzadeh,~\IEEEmembership{Member,~IEEE,}
        Jarkko~Kaleva,~\IEEEmembership{Member,~IEEE,}
        Nuutti~Tervo,~\IEEEmembership{Member,~IEEE,}
        Harri~Pennanen,~\IEEEmembership{Member,~IEEE,}
        Antti~Tölli,~\IEEEmembership{Senior Member,~IEEE,}
        and~Matti~Latva-aho,~\IEEEmembership{Senior Member,~IEEE.}
\thanks{Mohammad Majidzadeh, Nutti Tervo, Harri Pennanen, Antti Tölli, and Matti Latva-aho are with Centre for Wireless Communications (CWC), University of Oulu, Oulu, Finland (email: \{mohammad.majidzadeh, nuutti.tervo, harri.pennanen, antti.tolli, matti.latva-aho\}@oulu.fi). Jarkko Kaleva is with Solmu Technologies, Oulu, Finland, (email: jarkko.kaleva@gmail.com).

This research has been financially supported by Academy of Finland 6Genesis Flagship (grant no 318927) and Academy of Finland professor project (grant no 327035).}
}

\maketitle

\begin{abstract}

To satisfy the capacity requirements of future mobile systems, under-utilized millimeter wave frequencies can be efficiently exploited by employing massive MIMO technology with highly directive beamforming.
Hybrid analog-digital beamforming has been recognised as a promising approach for large-scale MIMO implementations with a reduced number of costly and power-hungry RF chains.
In comparison to fully connected architecture, hybrid beamforming (HBF) with partially connected RF architecture is particularly appealing for the practical implementation due to less complex RF power division and combining networks.
In this paper, we first formulate single- and multi-user rate maximization problems as weighted minimum mean square error (WMMSE) and derive solutions for hybrid beamformers using alternating optimization.
The algorithms are designed for the full-array- and sub-array-based processing strategies of partially connected HBF architecture.
In addition to the rate maximizing WMMSE solutions, we propose lower complexity sub-array-based zero-forcing algorithms.
The performance of the proposed algorithms is evaluated in two different channel models, i.e., a simple geometric model and a realistic statistical millimeter wave model known as NYUSIM.
The performance results of the WMMSE HBF algorithms are meant to reveal the potential of partially connected HBF and serve as upper bounds for lower complexity methods.
Numerical results imply that properly designed partially connected HBF has the potential to provide an good compromise between hardware complexity and system performance in comparison to fully digital beamforming.

\end{abstract}

\begin{IEEEkeywords}
hybrid analog-digital precoding, large scale MIMO, rate maximization, millimeter wave communications.
\end{IEEEkeywords}

\IEEEpeerreviewmaketitle

\section{Introduction}

\IEEEPARstart{I}{n} order to meet the high capacity and performance requirements of 5th generation (5G) mobile systems and beyond, new spectrum needs to be acquired at higher frequencies \cite{RappaportSMZAWWSSG, RohSPLLKCCA}.
Massive multiple-input multiple-output (MIMO) has been recognized as a promising approach to efficiently exploit the vast spectral resources available at millimeter waves, and freeing the great potential of millimeter wave (mm-wave) communications \cite{LarssonETM, MolischRHLNLH}.
Massive MIMO technology enables highly directive beamforming by utilizing a large amount of transmit (and/or receive) antennas \cite{LarssonETM}.
High beamforming gain is needed for supporting reasonable cell sizes by compensating severe free-space path loss occurring in mm-wave frequencies \cite{RanganRE}.
In addition, massive MIMO can provide spectrally efficient communications with high data rates through (multi-user) spatial multiplexing \cite{LarssonETM, RappaportSMZAWWSSG}.
Implementing massive MIMO using conventional digital beamforming is complex and hardware demanding since one radio frequency (RF) chain per antenna is needed \cite{SohrabiY}.
This is a costly and power consuming requirement due to the large number of needed RF components, such as wideband digital-to-analog converters (DACs) \cite{MolischRHLNLH, HanIXRl}.
In terms of hardware complexity, analog beamforming is a more feasible approach supporting single-stream transmission and requiring only a single RF chain regardless of the number of antennas \cite{RanganRE}.
However, analog beamforming cannot exploit the full potential of massive MIMO since spectrally efficient multi-stream transmission is not supported \cite{RanganRE}.
In this respect, hybrid analog-digital beamforming (HBF) is considered as a promising solution for implementing massive MIMO and providing a compromise between hardware complexity and spectral efficiency \cite{MolischRHLNLH, SohrabiY}.
The hybrid architecture splits the whole beamforming process into digital and analog parts enabling multi-stream transmissions with a reduced number of RF chains \cite{MolischRHLNLH, SohrabiY}.

\begin{figure}
	\begin{subfigure}{0.24\textwidth}
		\centering
		\includegraphics[width=.98\linewidth]{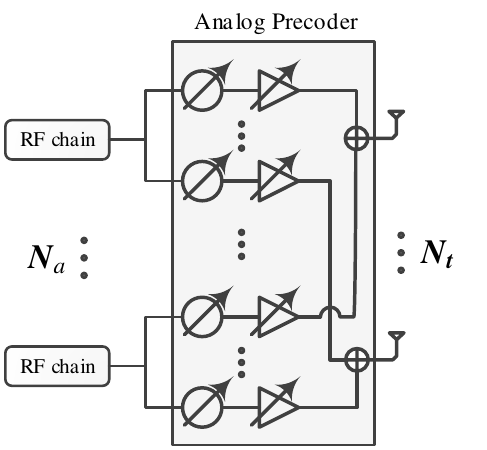}
		\caption{}
		\label{zFullyConnected}
	\end{subfigure}
	\begin{subfigure}{0.24\textwidth}
		\centering
		\includegraphics[width=.88\linewidth]{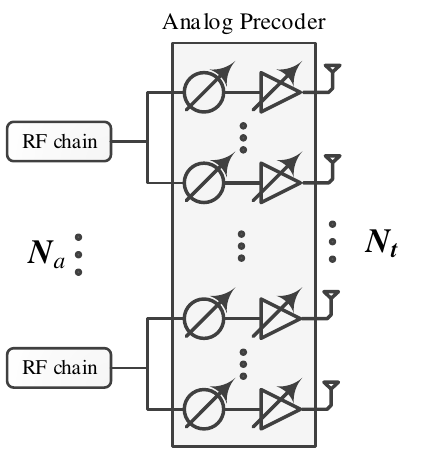}
		\caption{}
		\label{zPartiallyConnected}
	\end{subfigure}
	\caption{ (a) Fully connected and (b) Partially connected HBF architectures.}
	\label{zFullyPartiallyConnected}
\end{figure}

Most of the state-of-the-art hybrid methods are based on fully connected RF architecture  \cite{ShiH, LiN, UtschickSJL, WuLY, ComaUC, KimL, NguyenLLH, CongLZ, LinCZZB, RenWQL, HanifYBSS, HanifYBSSLow, NohKSL, YuZZ, LiWHG, IwanowBU, LiWYZY}, in which each RF chain is connected to all antennas as depicted in Fig. \ref{zFullyPartiallyConnected}.
However, due to very challenging and lossy RF signal division and combining processes, fully connected methods are of higher complexity and consume more power, hence over demanding to be implemented in systems with large antenna arrays.
A more practical solution is partially connected RF architecture, in which each RF chain is connected only to one subarray of antennas \cite{LiWYZY}.
Partially connected RF architecture itself can be divided into two categories, i.e., full array-based and subarray-based processing designs\cite{MajidzadehEuCNC, MajidzadehPIMRC, MajidzadehSPAWC}.
In full array-based processing, all data streams are conveyed to all subarrays.
Thus, each stream is transmitted via its corresponding beam which is generated by all subarrays.
This implies that full beamforming gain is potentially available.
However, the directions of different beams are interdependent due to the partial connectivity.
This sets restrictions to the beam generation and leads to suboptimal beam directions.
In subarray-based processing, each subarray transmits only a single data stream i.e., each data stream is conveyed to only one RF chain.
This leads to a more efficient and flexible beam design process.
However, the beamforming gain is limited by the number of antennas per subarray.

HBF has gained a lot of research interest in recent years.
As presented in a recent survey paper \cite{MolischRHLNLH}, HBF research can be divided into different categories based on, for example, the used system model, the level of channel knowledge and the wideness of bandwidth.
Each research category is important, and the corresponding research has its own merits increasing the knowledge on HBF.
This paper focuses on instantaneous CSI-based HBF for single- and multi-user settings in frequency flat channels.
Due to idealistic assumptions, the results in this paper can be considered as upper bounding benchmarks for more practical HBF designs.
In the literature, numerous fully and partially connected HBF approaches have been developed for single- and multi-user MIMO systems.
In the following, the main prior works are introduced.

Other research works in the literature have used different criteria and methodologies than ours and the approaches to solve the problem are quite diverse. In the following, we will discuss some examples that are somehow related to our work, however, either the problem formulation or the approaches to tackle the problems are different from ours.
For digital SU-MIMO systems, the SVD based beamforming is well known to be the optimal solution, however, it is not applicable in HBF.
Hence, other approaches have been used to solve rate maximization problems.
Authors in \cite{ChenTLHW} proposed an iterative alternating optimization based algorithm where the optimal digital precoder is approximated by a feasible hybrid precoder based on the minimum mean square error (MMSE) principle.
In \cite{MirzaANS}, a HBF solution for the considered rate maximization problem by approximating it as a single-stream beamforming problem with per-antenna power constraints is developed.
A HBF design with phase shifter selection was studied for large antenna arrays in \cite{PayamiGD}.
Authors in \cite{SohrabiYOFDM} proposed an OFDM based heuristic HBF algorithm for both fully connected and partially connected architectures.
A method using directivity of the channel is proposed for SU-MIMO systems in \cite{RaghavanSCSKL}.
In \cite{JinZCX}, a HBF design based on matrix factorization with finite-alphabet inputs is proposed.
An optimal transmit HBF method for  partially connected massive MISO system with RF chain and per-antenna power constraints is presented in \cite{CaoOS}.
Authors in \cite{VlachosKT} designed an energy efficient transmitter with low resolution DACs for partially connected architecture. However, this paper tries to tackle the rate maximization problem with a WMMSE based alternating optimization approach. Moreover, a lower complexity ZF based algorithm is proposed.

In multi-user massive MISO systems, MMSE based algorithms have been used to tackle the nonconvex sum rate maximization problems. Moreover, lower complexity transmission algorithms, such as zero forcing (ZF) schemes are of practical interest since they may provide relatively good performance with low computational complexity.
Authors in \cite{XingyuLZZ} have used WMMSE formulation to solve the HBF optimization problem for partially connected architecture in a way which is close to our formulation, however, the approaches to solve the problems are quite different from our case. They have used unit modulus constraint for their analog beamformer and two different approaches named element iteration (IT) and manifold optimization (MO) to solve the problem which are different from our methods.
In \cite{LinCZZB}, the MMSE criterion has been considered as the HBF optimization problem, where the HBF design problem is decomposed into two sub-problems i.e., precoding and combining problems, and they are solved in a unified manner but for fully connected architecture.
A ZF-based HBF approach was developed for mm-wave massive MU-MIMO communications in \cite{PayamiGD}.
In \cite{YuQFJZ} a two-stage HBF method is proposed. First, the lower bound of sum rate is derived and then an iterative algorithm is designed to improve the performance.
A hybrid precoding architecture is presented in \cite{CastellanosRRKLLP} that allows both amplitude and phase control at low complexity and cost to allow the implementation of the zero-forcing structure.
An analysis framework is presented in \cite{SongKC} to compare fully and partially connected architectures with one stream per subarray.
In \cite{LiHSWM}, a block descent algorithm is used to solve a conditional average net sum rate maximizing problem via an equivalent weighted average mean square error minimization (WAMMSE) problem. 

Most of the HBF works in the literature use only phase shifting in the analog beamforming process.
However, employing analog amplitude control in addition to phase shifting can be assumed to be also a feasible assumption \cite{MajidzadehEuCNC, MajidzadehPIMRC, MajidzadehSPAWC, RusuMGH, CastellanosRRKLLP}.
Even though this will somewhat increase the hardware complexity, the design of hybrid beamforming weights becomes more flexible and efficient leading to improved performance.
As an example, fully connected HBF with only analog phase shifting can obtain the same performance as fully digital beamforming if the number of RF chains is twice the number of data streams \cite{SohrabiY}.
In comparison, HBF with both analog phase shifting and amplitude control can provide equal performance compared to fully digital beamforming when the number of RF chains and streams is the same.
Only few algorithms have been proposed for partially connected HBF employing both analog amplitude and phase control \cite{MajidzadehEuCNC, RusuMGH}.
Authors in \cite{MajidzadehPIMRC} proposed an algorithm that minimizes the mean square error (MSE) between the optimal precoder and hybrid beamforming matrix.
In \cite{MajidzadehEuCNC}, several HBF algorithms were developed including singular value decomposition matching, iterative orthogonalization, and transmit-receive zero-forcing.
Different optimization-based hybrid methods were devised for frequency selective MIMO systems in \cite{RusuMGH}.
Due to the scarcity of algorithms and performance evaluations in the literature, partially connected HBF with both analog amplitude and phase control needs further  studying.
Performance  comparisons  between full array and subarray-based heuristic and optimization-based HBF strategies are of particular interest

\begin{figure*}
	\centering
	\includegraphics[width=.84 \linewidth]{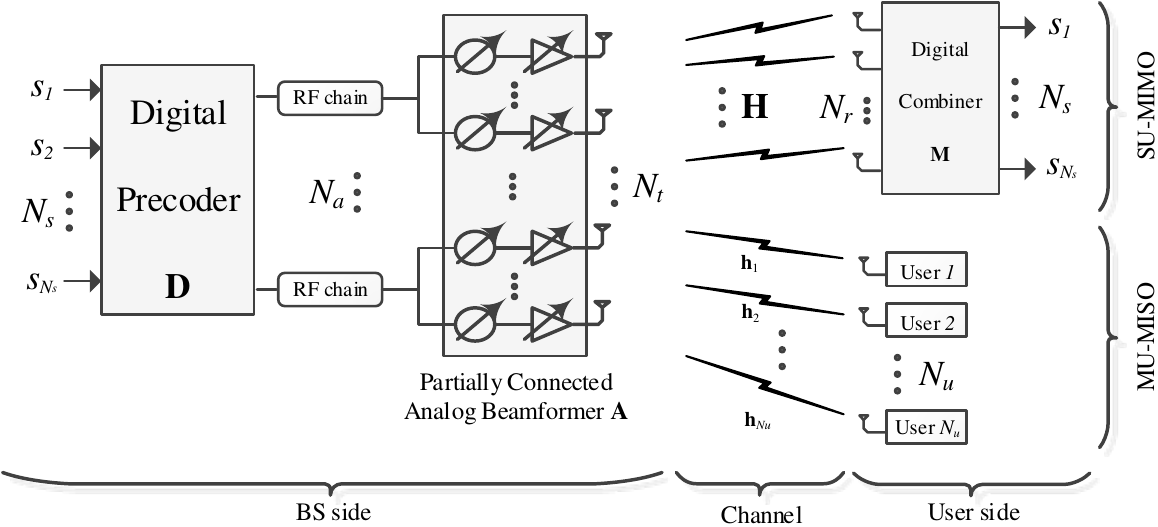}
	\caption{Overall SU-MIMO and MU-MIMSO system models}
	\label{systemmodel}
\end{figure*}

In this paper, several optimization-based and heuristic HBF algorithms are proposed for SU-MIMO and MU-MISO systems with partially  connected  RF  architecture.
Algorithms are designed for both full array-and  subarray-based  HBF processing strategies.
In addition to phase shifting, amplitude control is also employed in the analog part of the HBF process.
Inspired by the existing WMMSE algorithms in the digital beamforming literature such as \cite{KalevaTJ}, we first reformulate the considered SU and MU rate maximization problems as weighted MSE minimization problems.
These problems are solved by alternately optimizing between the receive combiner, the digital precoder and the analog beamformer until the objective value converges.
This is to find solutions that can perform as upper bounds for systems with similar configurations and to avoid fully connected architecture due to its high complexity.
The proposed hybrid WMMSE algorithms are designed for both full array-and subarray-based HBF processing strategies.
Since the original problems are non-convex, the optimality of the solutions can not be guaranteed.
To the best of our knowledge, this kind of WMMSE algorithms haven't been proposed in the literature for the considered hybrid systems.
In addition to rate maximizing WMMSE methods, we propose lower complexity  subarray based ZF algorithms.
An iterative subarray-based SU transmit-receive ZF algorithm aims at canceling inter-stream interference using ZF beamforming at both transmitter and receiver.
A simple subarray-based ZF scheme is also considered for MU-MISO.
Numerical simulations are conducted to evaluate the performance of the proposed HBF algorithms against fully digital and analog (one stream) beamforming solutions in two different environments. One is a simple geometric uniform linear array (ULA) and the other is a more realistic NYUSIM channel model.

The following notations are used throughout this paper.
Uppercase boldface and lowercase boldface characters denote matrices and vectors, respectively.
${\rm diag}(.)$ denotes a diagonal matrix of its arguments.
The superscripts $(.)^{H}$, $(.)^{T}$, and $(.)^{\ast}$, indicate Hermitian, transpose, and complex conjugate, respectively.
The matrix ${\bf I}_M$ denotes an $M \times M$ identity matrix.
$|.|$ and ${\rm tr}(.)$ are used to represent determinant and trace of a matrix, respectively.

The remainder of the paper is organized as follows.
In Section \ref{SystemModel}, the employed system model of SU-MIMO and MU-MISO scenarios are described.
Section \ref{ProblemFormulation} presents the formulation of the problems in both scenarios.
Section \ref{HBFAlgorithmsforSUMIMO} and \ref{HBFAlgorithmsforMUMISO} investigate the proposed algorithms for SU-MIMO and MU-MISO scenarios, respectively.
The simulation results are presented and discussed in Section \ref{SimulationResults}.
Finally, Section \ref{Conclusion} concludes the paper.

\section{System Model}
\label{SystemModel}

This section introduces SU-MIMO and MU-MISO system models assuming partially connected RF architecture at the transmitter side.
These models are depicted in Fig. \ref{systemmodel}.
First, the main system details are described, then signal models are presented for both systems separately.

Consider the systems to be in the downlink mode and have identical settings at the base station (BS) side.
The number of receive antennas/users is assumed to be considerably smaller than the number of transmit antennas. Moreover, the analog beams at the users are statistically fixed towards the serving BS. Hence, it can be assumed to be part of the channel information.
Thus, HBF is considered only at the BS and the receiver is assumed to be fully digital.
The BS is equipped with $N_t$ transmit antennas and a hybrid architecture with $N_a$ RF chains.
In order to find solutions that can perform as upper bounds for partially connected HBF systems with similar configuration, it is assumed that channel state information (CSI) is available at the transmitter as well as the receiver.
In the single-user case, the user has $N_r$ receive antennas and the number of data streams is $N_s$ which is assumed to be equal to the number of RF chains.
In the multi-user case, $N_u$ single antenna users are served by the BS and the number of data streams $N_s$ is equal to the number of users and the number of RF chains.
Moreover, partially connected architecture is used in which each RF chain is connected to only one subarray of the antennas.
The transmit antenna array is partitioned into $N_a$ subarrays each with $n = N_t/N_a$ antennas.
In the overall hybrid architecture at the BS, amplitude control is employed in addition to phase shifting in the analog domain.

\begin{figure}
        \begin{subfigure}{0.24\textwidth}
        \centering
                \includegraphics[width=.98\linewidth]{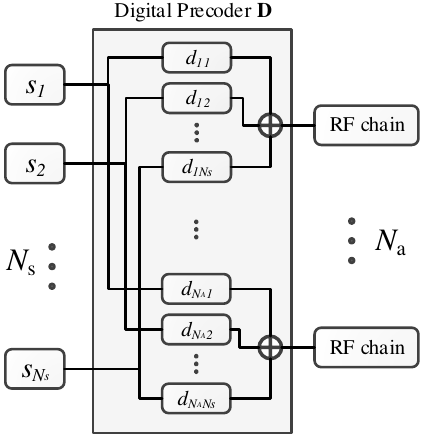}
                \caption{}
                \label{zFullarraybased}
        \end{subfigure}
        \begin{subfigure}{0.24\textwidth}
        \centering
                \includegraphics[width=.85\linewidth]{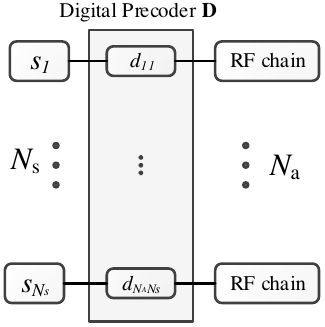}
                \caption{}
                \label{zSubarraybased}
        \end{subfigure}
        \caption{ (a) Full array- and (b) subarray-based processing strategies for partially connected HBF.}
       \label{zFullSubArray}
\end{figure}

The HBF architecture consists of a digital precoder ${\bf D}$ and an analog beamformer ${\bf A}$.
Based on the digital precoder structure, two different designs, i.e., full array-based and subarray-based processing methods can be considered as depicted in Fig. \ref{zFullSubArray}.
In the case of full array-based processing, all data streams are connected to all RF chains and the digital precoder ${\bf D} \in \mathbb{C}^{N_a \times N_s}$ is given by
\begin{equation}
\begin{split}
{\bf D}
& =\left(
\begin{array}{cccc}
d_{11} & d_{12} & \ldots & d_{1N_s}\\
d_{21} & d_{22} & \ldots & d_{2N_s}\\
\vdots & \vdots & \ddots & \vdots\\
d_{N_a1} & d_{N_a2} & \ldots & d_{N_aN_s}
\end{array} \right)
=\left(
\begin{array}{cccc}
{\bf d}_{1}\\
{\bf d}_{2}\\
\vdots\\
{\bf d}_{N_a}
\end{array} \right)\\
& \hspace{1mm} = \left( \hspace{3mm}
\begin{array}{cccc}
{\bf \bar d}_{1} & \hspace{4mm} {\bf \bar d}_{2} & \hspace{1mm} \ldots & \hspace{2mm} {\bf \bar d}_{N_s}\\
\end{array} \hspace{2mm} \right)
\end{split}
\label{VDfull}
\end{equation}
where ${\bf d}_{i}  = (\,d_{i1}\, d_{i2} \,\ldots \,d_{i N_s}\,)$ and ${\bf \bar d}_{j}  = (\,d_{1j}\, d_{2j} \,\ldots \,d_{N_a j}\,)^T$ are the $i$th and the $j$th row and column vectors of the digital precoder, corresponding to the $i$th and the $j$th subarray and stream, respectively.
In the case of subarray-based processing, where each data stream is connected to only one RF chain (i.e., $N_a=N_s$), the digital precoder becomes a diagonal matrix ${\bf D} = {\rm diag}\,(\,d_{11}\, d_{22}\, \ldots\, d_{N_s N_s}\,)$.
To simplify more, the digital weights can be directly incorporated into the analog beamformer amplitudes and phases of the corresponding subarray.
Thus, the digital precoder can be normalized to be identity matrix which only routes data streams to the RF chains.
The analog beamformer ${\bf A} \in \mathbb{C}^{N_t \times N_a}$ can be expressed as
\begin{equation}
\begin{split}
{\bf A}
=\left(
\begin{array}{cccc}
{\bf a}_1 & {\bf 0} &  \ldots & {\bf 0}\\
{\bf 0} & {\bf a}_2 & \ldots & {\bf 0}\\
\vdots & \vdots & \ddots & \vdots\\
{\bf 0} & {\bf 0} & \ldots & {\bf a}_{N_a}
\end{array} \right)
\end{split}
\end{equation}
where ${\bf a}_i \in {\mathbb C}^{n \times 1}$ is the analog RF beamformer of the $i$th subarray, and ${\bf 0} \in {\mathbb C}^{n \times 1}$ is a zero vector.

\subsection{Signal Model for SU-MIMO}

In the SU-MIMO scenario, the estimated signal vector at the user is given by
\begin{equation}
\begin{split}
\hat{{\bf s}} & = {\bf M}^H {\bf H} {\bf A} {\bf D} {\bf s} + {\bf M}^H {\bf z}
\end{split}
\label{receivedsignal}
\end{equation}
where ${\bf H} \in {\mathbb C}^{N_r \times N_t}$ denote the channel matrix, ${\bf s} = (s_1, s_2, \ldots, s_{N_s})^T \in {\mathbb C}^{N_s \times 1}$ is the vector of data symbols with ${\mathbb E}[{\bf ss}^H]={\bf I}_{N_s}$, ${\bf z} \thicksim \mathcal{CN}({\bf 0},N_0{\bf I}_{N_r})$ stands for additive white Gaussian noise, and ${\bf M} = (\,{\bf m}_1 \; {\bf m}_2 \; \ldots \; {\bf m}_{N_s}\,) \in {\mathbb C}^{N_r \times N_s}$ denotes the digital receive beamformer.
The vector ${\bf m}_i  \in {\mathbb C}^{N_r \times 1}$ is the $i$th receive beamformer of the corresponding spatial data stream.

The rate of stream $i$ can be written as

\begin{equation}
R_i = \log_2\left(1 + {|{\bf m}_i^H {\bf H} {\bf v}_i|^2 \over N_0 + \sum\limits_{\substack{l=1, l \neq i}}^{N_s} |{\bf m}_i^H {\bf H} {\bf v}_l|^2} \right)
\label{subarrayrate}
\end{equation}
where
${\bf v}_{i} = ({\bf a}_1 d_{1i},~ {\bf a}_2 d_{2i},~ \ldots,~ {\bf a}_{N_a} d_{N_ai})^T \in {\mathbb C}^{N_t \times 1}$
is the overall hybrid precoder of stream $i$.

\subsection{Signal Model for MU-MISO}

In the MU-MISO scenario, the received signal of the $k$th user can be modeled as
\begin{equation}
\begin{split}
y_k
&= {\bf h}_{k}^{H} {\bf A} {\bf \bar d}_{k} s_{k} + {\bf h}_{k}^{H} \sum\limits_{\substack{i=1\\ i \neq k}}^{N_u} {\bf A} {\bf \bar d}_{i} s_{i} + z_k\\
\end{split}
\label{kthreceivedsignal}
\end{equation}
where ${\bf h}_{k} \in  {\mathbb C}^{N_t \times 1}$ is the channel vector from the transmitter to the $k$th user,
${\bf \bar d}_{k} \in  {\mathbb C}^{N_a \times 1}$ is the digital precoder corresponding to the $k$th user,
$s_{k}$ is the data symbol of the user $k$, and $z_k \thicksim \mathcal{CN}(0,N_0)$ is the additive white Gaussian noise of the $k$th user.

The rate expression for user $k$ can be written as
\begin{equation}
R_k = \log_2\left(1 + {|m_k {\bf h}_{k}^{H} {\bf A} {\bf \bar d}_{k}|^2 \over N_0 + \sum\limits_{\substack{l=1, l \neq k}}^{N_u} |m_k {\bf h}_{k}^{H} {\bf A} {\bf \bar d}_{l}|^2}\right)
\label{subarrayrate}
\end{equation}
where $m_k$ is the digital receiver of user $k$ which only scales and shifts the phase.

\section{Problem Formulation}
\label{ProblemFormulation}

Our objective is to maximize the data rate of the system while satisfying the total transmission power constraint at the BS.
In the following, the rate maximization problem is formulated for SU-MIMO and MU-MISO systems.
The optimization problems are equivalently reformulated as weighted MSE minimization that can be solved using iterative alternating optimization efficiently. 

\subsection{Rate Maximization for SU-MIMO}

The optimization objective in the considered SU-MIMO system is to maximize the rate of the user while satisfying the maximum transmission power constraint.
This rate maximization problem is expressed as
\begin{equation}
\begin{split}
& \operatorname*{maximize}_{{\bf A}, {\bf D}, {\bf M}}
\sum_{i=1}^{N_s} R_i\\
& \hspace{-5mm}  {\rm s. t.} \hspace{.3cm} {\rm tr}\,({\bf A} {\bf D} {\bf D}^{H} {\bf A}^{H}) \leq  P
\end{split}
\label{WSRMaxproblem}
\end{equation}
where $P$ is the maximum transmission power at the BS.
Solving (\ref{WSRMaxproblem}) requires digital precoder $\bf D$, analog beamformer $\bf A$, and receive beamformer $\bf M$ to be optimized jointly.
However, this joint optimization problem is non-convex and cannot be optimally solved in its current form.
In the Section \ref{ProblemFormulation}, we reformulate (\ref{WSRMaxproblem}) as a WMMSE optimization problem.
This problem is still non-convex but it can be solved by using an iterative alternating optimization method, however, the solution is not guaranteed to be globally optimal.

The error matrix at the output of the receive beamformer is given by
\begin{equation}
\begin{split}
{\bf E}
& ={\mathbb E}\left[ \left( {\bf s} - {\bf M}^H {\bf y} \right) \left( {\bf s} - {\bf M}^H {\bf y} \right)^H \right]\\
& = {\bf I} - {\bf  M}^H {\bf H} {\bf A} {\bf D} - {\bf D}^H {\bf A}^H {\bf H}^H {\bf M} + N_0 {\bf M}^H {\bf M}\\
& \hspace{7mm} + {\bf M}^H {\bf H} {\bf A} {\bf D} {\bf D}^H {\bf A}^H {\bf H}^H {\bf M}.
\end{split}
\label{E}
\end{equation}
The well known rate optimal digital MMSE receive beamformer can be derived as
\begin{equation}
\begin{split}
{\bf M}
& = ({\bf H} {\bf A} {\bf D} {\bf D}^H {\bf A}^H {\bf H}^H + N_0 {\bf I}_{N_r})^{-1} {\bf H} {\bf A} {\bf D}.
\end{split}
\label{MMSEcombiner}
\end{equation}
The error matrix after applying the receive beamformer is given by
\begin{equation}
\begin{split}
{\bf E}
& = ({\bf I} + \frac{1}{N_0}{\bf H} {\bf A} {\bf D} {\bf D}^H {\bf A}^H {\bf H}^H )^{-1}
\end{split}
\end{equation}

It can be shown that $R = \log_2 |{\bf E}^{-1}|$.
By successive approximation of the objective function, the rate maximization problem for fully digital beamforming can be iteratively solved via WMMSE optimization \cite{ChristensenACC}.
Similarly, for fixed approximation coefficients (weights), our corresponding HBF optimization problem can be written as
\begin{equation}
\begin{split}
& \operatorname*{minimize}_{{\bf A}, {\bf D}, {\bf M}}
{\rm tr}\, ({\bf W} {\bf E})\\
& \hspace{-5mm} {\rm s. t.} \hspace{.3cm} {\rm tr}\,({\bf A} {\bf D} {\bf D}^{H} {\bf A}^{H}) \leq  P
\end{split}
\label{WMSEminproblemMat}
\end{equation}
where ${\bf W} = {\rm diag}\,(\,w_1\, w_2\, \ldots\, w_{N_s}\,)$ is the weight matrix,
\begin{equation}
w_i = e_i^{-1}
\label{weights}
\end{equation}
is the weight of stream $i$,
and
\begin{equation}
\begin{split}
e_i
& =  \left( 1 + {\bf v}_{i}^H {\bf H}^H \left( {\bf H}  \sum \limits_{\substack{l=1 \\ l\neq i}}^{N_s} {\bf v}_l  
{\bf v}_{l}^H {\bf H}^H + N_0 {\bf I} \right)^{-1} {\bf H} {\bf v}_i \right)^{-1}
\end{split}
\label{eiafter}
\end{equation}
is the error term corresponding to stream $i$.
Since
$\sum_{i=1}^{N_s} R_i = \sum_{i=1}^{N_s} \log_2 |e_i^{-1}|$,
we can write the WMMSE problem as
\begin{equation}
\begin{split}
& \operatorname*{minimize}_{\{{\bf a}_i\}, \{{\bf d}_i\}, \{{\bf m}_i\}}
\sum_{i=1}^{N_s} w_i e_i\\
& \hspace{-5mm} {\rm s. t.} \hspace{.3cm} \sum_{i=1}^{N_s} {\rm tr}\,({\bf a}_i {\bf d}_i {\bf d}_i^{H} {\bf a}_i^{H}) \leq  P.
\end{split}
\label{WMSEminproblemVec}
\end{equation}
In Section \ref{HBFAlgorithmsforSUMIMO}, we propose HBF algorithms to suboptimally solve the non-convex WMMSE problem using alternating optimization over the receive beamformer, digital precoder, and analog beamformer.

\subsection{Sum Rate Maximization for MU-MISO}

The MU-MISO optimization problem aims at maximizing the sum rate of the users with a maximum transmission power constraint.
The sum rate maximization problem ca be written as
\begin{equation}
\begin{split}
& \operatorname*{maximize}_{{\bf A}, \{{\bf \bar d}_k\}}
\sum_{k=1}^{N_u} R_k\\
& \hspace{-5mm}  {\rm s. t.} \hspace{.3cm} \sum_{k=1}^{N_u} {\rm tr}\,({\bf A} {\bf \bar d}_k {\bf \bar d}_k^{H} {\bf A}^{H}) \leq  P.
\end{split}
\label{SRMaxProblemMUMISO}
\end{equation}
Joint optimization of the digital precoders $\{{\bf \bar d}_k\}$ and the analog beamformer $\bf A$ is non-convex and cannot be optimally solved in its current form.
In the following, we reformulate (\ref{SRMaxProblemMUMISO}) as a WMMSE problem.
Similar to SU-MIMO, this problem is still non-convex but it can be solved by using iterative alternating optimization, however, the solution is not guaranteed to be globally optimal.

The error term at the $k$th user is given by
\begin{equation}
\begin{split}
e_k
& ={\mathbb E}\left[ \left( s_k - m_k y_k \right) \left( s_k - m_k y_k \right)^H \right]\\
& = 1 - m_k {\bf h}_k^H {\bf A} {\bf \bar d}_k  - {\bf \bar d}_k^H {\bf A}^H {\bf h}_k m_k^{\ast}\\
& \hspace{7mm} +  m_k {\bf h}_k^H {\bf A} {\bf D} {\bf D}^H {\bf A}^H {\bf h}_k m_k^{\ast} + m_k m_k^{\ast} N_0
\end{split}
\label{ek}
\end{equation}

The rate optimal MMSE receiver of user $k$ can be derived as
\begin{equation}
\begin{split}
m_k
& = {\bf \bar d}_k^H {\bf A}^H {\bf h}_k ({\bf h}_k^H {\bf A} {\bf D} {\bf D}^H {\bf A}^H {\bf h}_k + N_0)^{-1}.
\end{split}
\label{MMSEUserReceiver}
\end{equation}
After applying the MMSE receiver to (\ref{ek}), the error term at the $k$th user is given by
\begin{equation}
\begin{split}
e_k
& = 1 - {\bf \bar d}_k^H {\bf A}^H {\bf h}_k ({\bf h}_k^H {\bf A} {\bf D} {\bf D}^H {\bf A}^H {\bf h}_k + N_0)^{-1} {\bf h}_k^H {\bf A} {\bf \bar d}_k
\end{split}
\label{ekafter}
\end{equation}

Similar to the SU-MIMO case, $R_k = \log_2 |e_k^{-1}|$.
For fixed approximation coefficients (weights), our corresponding HBF optimization problem can be written as
\begin{equation}
\begin{split}
& \operatorname*{minimize}_{{\bf A}, \{{\bf \bar d}_k\}}
\sum_{k=1}^{N_u} w_k e_k\\
& \hspace{-5mm}  {\rm s. t.} \hspace{.3cm} \sum_{k=1}^{N_u} {\rm tr}\,({\bf A} {\bf \bar d}_k {\bf \bar d}_k^{H} {\bf A}^{H}) \leq  P
\end{split}
\label{WMSEminproblemMUMISO}
\end{equation}
where
\begin{equation}
w_k = e_k^{-1}
\label{wk}
\end{equation}
is the weight of user $k$.
In Section \ref{HBFAlgorithmsforMUMISO}, we propose HBF algorithms to suboptimally solve the non-convex WMMSE problem using alternating optimization over the digital precoder and the analog beamformer.

\section{HBF Algorithms for SU-MIMO}
\label{HBFAlgorithmsforSUMIMO}

In this section, three HBF algorithms are proposed for SU-MIMO systems with partially connected RF architecture.
One algorithm is developed for full-array based hybrid design and two for sub-array-based processing strategy.
The derivations of these algorithms are presented in the following subsections.

\subsection{Full Array-Based Hybrid WMMSE}

The aim of this algorithm is to solve the rate maximization problem via an equivalent reformulation of weighted MSE minimization.
By applying an iterative alternating optimization method, the WMMSE problem can be solved so that the objective value converges with a desired accuracy.
This algorithm consists of three main steps.
First, (\ref{WMSEminproblemMat}) is solved with respect to the receive beamformer ${\bf M}$ while the digital and analog beamformers are fixed.
Then, the analog beamformer ${\bf A}$ and the receive beamformer ${\bf M}$ are kept fixed and (\ref{WMSEminproblemMat}) is solved for the digital precoder ${\bf D}$.
Last step is to optimize the analog beamformer ${\bf A}$ while keeping the other two variables fixed.
In the following, the hybrid WMMSE algorithm is described in detail.

\begin{algorithm}[t]
  \caption{Full Array-Based Hybrid WMMSE Algorithm}\label{FullarrayHybridWMMSEalgorithm}
  \begin{algorithmic}[1]
      \State Set iteration number $n=0$ and initialize ${\bf D}^{n}$ and ${\bf A}^{n}$.
      \Repeat
      \State Update $n=n+1$.
      \State Solve (\ref{MMSEcombiner}) for ${\bf M}^n$ while ${\bf D}^{n-1}$ and ${\bf A}^{n-1}$ are fixed.
      \State Update ${\bf W}^{n}$ from (\ref{weights}), (\ref{eiful}) given ${\bf D}^{n-1}$, ${\bf A}^{n-1}$, and $\bf M^{n}$.
      \State Solve (\ref{DMMSE}) for ${\bf D}^{n}$ while ${\bf M}^{n}$ and ${\bf A}^{n-1}$ are fixed.
      \State Solve (\ref{aiful}) for $\{{\bf a}_{i}^{n}\}$ while ${\bf M}^{n}$ and ${\bf D}^{n}$ are fixed.
      \Until{desired level of convergence}
  \end{algorithmic}
\end{algorithm}

Problem (\ref{WMSEminproblemMat}) is convex with respect to the receive beamformer $\bf M$. The Lagrangian expression of (\ref{WMSEminproblemMat}) is given by
\begin{equation}
\mathcal{L} = {\rm tr}\, ({\bf W} {\bf E}) + \alpha ({\rm tr}\,({\bf A} {\bf D} {\bf D}^{H} {\bf A}^{H}) - P)
\label{LagFulM}
\end{equation}
where $\alpha$ is the Lagrange multiplier. The first order optimality condition with fixed digital and analog beamformers yields the MMSE receive beamformer $\bf M$ in (\ref{MMSEcombiner}).
The next step is to solve (\ref{WMSEminproblemMat}) for ${\bf D}$.
The resulting expression from the first order optimality condition is
\begin{equation}				
{\bf D} = ({\bf A}^H {\bf H}^H {\bf M} {\bf W} {\bf M}^H {\bf H} {\bf A} + \alpha {\bf A}^H {\bf A})^{-1} {\bf A}^H {\bf H}^H {\bf M} {\bf W}
\label{DMMSE}
\end{equation}
where $\alpha \geq 0$ is chosen such that the transmit power constraint is satisfied. If $\alpha = 0$ satisfies the transmit power constraint, the solution is ready. Otherwise $\alpha > 0$ can be found using one dimensional search techniques such as the bisection method.
In the next step, we fix the digital precoder and the receive beamformer and solve (\ref{WMSEminproblemVec}) with respect to the analog beamformers for different subarrays.
Rewriting the received signal as
\begin{equation}
\begin{split}
{\bf y} & = \sum_{j=1}^{N_a} {\bf H}_j {\bf a}_j {\bf d}_{j} {\bf s} + {\bf z}
\end{split}
\label{receivedsignalRE}
\end{equation}
yields the error term corresponding with stream $i$ as
\begin{equation}
\begin{split}
e_i
&= 1 - {\bf  m}_i^H \sum_{j=1}^{N_a} {\bf H}_j {\bf a}_j d_{ji} - \sum_{j=1}^{N_a} d_{ji}^{\ast} {\bf a}_j^H {\bf H}_j^H {\bf m}_i\\
&\hspace{7mm} + {\bf m}_i^H \sum_{j=1}^{N_a} {\bf H}_j {\bf a}_j {\bf d}_j \sum_{l=1}^{N_a} {\bf d}_{l}^H {\bf a}_l^H {\bf H}_l^H {\bf m}_i + N_0 {\bf m}_i^H {\bf m}_i.
\end{split}
\label{eiful}
\end{equation}
where sub-channel ${\bf H}_j \in {\mathbb C}^{N_r \times n}$ is the matrix of complex channel gains between transmit antennas of the $j$th subarray and $N_r$ receive antennas. The channel matrix ${\bf H}$ can be written as ${\bf H} = (\,{\bf H}_1 \; {\bf H}_2 \; \ldots \; {\bf H}_{N_a}\,)$.
Consequently, the Lagrangian corresponding to (\ref{WMSEminproblemVec}) can be expressed as
\begin{equation}
\begin{split}
\mathcal{L}
& = \sum_{i=1}^{N_s} w_i e_i + \alpha (\sum_{i=1}^{N_s} {\rm tr}\,({\bf a}_i {\bf d}_i {\bf d}_i^{H} {\bf a}_i^{H}) -  P).
\end{split}
\label{LagrangianSub}
\end{equation}
The first order optimality condition with respect to ${\bf a}_i$ yields (\ref{aiful})
in which $\alpha \geq 0$ is chosen via the bisection method such that the transmit power constraint is satisfied.
This procedure of alternating optimization continues until a desired level of convergence is achieved.
The proposed algorithm converges in terms of objective value since solving a convex problem at each step improves the objective value and the resulting MSE is lower bounded \cite{ShiRLH}.
The optimality of the solution cannot be guaranteed due to the non-convexity of the original problem.
Hence, the solution has to be treated as suboptimal unless otherwise proven.
Algorithm \ref{FullarrayHybridWMMSEalgorithm} summarizes the proposed full array-based hybrid WMMSE approach.

\begin{figure*}
\begin{equation}
{\bf a}_i = ({\bf H}_i^H {\bf M} {\bf W} {\bf M}^H {\bf H}_i + \alpha {\bf I}_n)^{-1} {\bf H}_i^H \bigg( \sum_{j=1}^{N_s} {\bf m}_j w_j d_{ij}^{\ast} - {\bf M} {\bf W} {\bf M}^H \sum \limits_{\substack{l=1 , l\neq i}}^{N_s} {\bf H}_l {\bf a}_l {\bf d}_l {\bf d}_i^H \bigg) ({\bf d}_i {\bf d}_i^H)^{-1}
\label{aiful}
\end{equation}
\end{figure*}

\subsection{Subarray-Based Hybrid WMMSE}

\begin{algorithm}[t]
  \caption{Subarray-Based Hybrid WMMSE Algorithm}\label{SubarrayHybridWMMSEalgorithm}
  \begin{algorithmic}[1]
      \State Set iteration number $n=0$ and initialize ${\bf A}^{n}$.
      \Repeat
      \State Update $n=n+1$.
      \State Solve (\ref{MMSEcombiner}) for ${\bf M}^{n}$ while ${\bf A}^{n-1}$ is fixed.
      \State Update ${\bf W}^{n}$ from (\ref{weights}) given ${\bf A}^{n-1}$, and ${\bf M}^{n}$.
      \State Solve (\ref{aisub}) for $\{{\bf a}_{i}^{n}\}$ while ${\bf M}^{n}$ is fixed.
      \Until{desired level of convergence}
  \end{algorithmic}
\end{algorithm}

The sub-array-based WMMSE method is a simplified version of Algorithm \ref{FullarrayHybridWMMSEalgorithm}, since the digital beamformer is fixed as an identity matrix.
Now the weighted MSE minimization problem in (\ref{WMSEminproblemMat}) can be solved by alternating between optimization of the receive beamformer ${\bf M}$ and the analog beamformer ${\bf A}$.
In the first step, the problem is solved with respect to the receive beamformer ${\bf M}$.
The first order optimality condition of (\ref{WMSEminproblemMat})
with respect to ${\bf M}$ yields the same MMSE receive beamformer as in (\ref{MMSEcombiner}).
Then, the problem is solved for the analog beamformer ${\bf A}$.
Using the stream specific MSE expressions, the corresponding Lagrangian is given by
\begin{equation}
\begin{split}
\mathcal{L} =
& \sum_{i=1}^{N_s} w_i e_i + \alpha (\sum_{i=1}^{N_s} {\rm tr}\,({\bf a}_i {\bf a}_i^{H}) -  P).
\end{split}
\label{LagrangianSub}
\end{equation}
The first order optimality condition of $\mathcal{L}$ with respect to each ${\bf a}_i$ yields
\begin{equation}
{\bf a}_j = ({\bf H}_j^H {\bf M} {\bf W} {\bf M}^H {\bf H}_j + \alpha {\bf I}_n)^{-1} {\bf H}_j^H {\bf m}_j w_j
\label{aisub}
\end{equation}
where $\alpha \geq 0$ is chosen via the bisection method while satisfying the transmit power constraint.
This alternating optimization procedure is repeated until a desired level of convergence is obtained.
The developed subarray-based hybrid WMMSE approach is summarized in Algorithm \ref{SubarrayHybridWMMSEalgorithm}.
The convergence and suboptimality properties of Algorithm \ref{FullarrayHybridWMMSEalgorithm} apply also for Algorithm \ref{SubarrayHybridWMMSEalgorithm}.

\subsection{Subarray-Based Transmit-Receive ZF}

\begin{algorithm}[t]
  \caption{Subarray-Based Transmit-Receive ZF Algorithm}\label{transmitreceiveZFalgorithm}
  \begin{algorithmic}[1]
      \State SVD: ${\bf H}={\bf \Gamma} {\bf \Sigma} {\bf \Lambda}^H$
      \State ${\bf M} \gets $ first $N_s$ columns of ${\bf \Gamma}$
      \Repeat
      \For{$i=1:N_s$\texttt{}}
      \State ${\bf V}_i \gets {\bf H}_i^{H} {\bf M} ({\bf M}^{H} {\bf H}_i {\bf H}_i^{H} {\bf M})^{-1}$
      \State Subarray $i$ entries of $i$th column of ${\bf {\bar V}} \gets$ normalized $i$th column of ${\bf V}_i$
      \EndFor
      \State ${\bf M}^{H} \gets ({\bf {\bar V}}^{H} {\bf H}^{ H} {\bf H}{\bf {\bar V}})^{-1} {\bf {\bar V}}^{H} {\bf H}^{H}$
      \Until{meeting \{$R$ threshold $|$ Max iterations\}}
      \State Calculate power allocations $\{P_i\}$, $i=1, 2, \ldots, N_s$ via WF using (\ref{waterfilling}), (\ref{eigenvalues})
      \State ${\bf V} \gets \sqrt{\bf P} {\bf {\bar V}}$
  \end{algorithmic}
\end{algorithm}

Transmit-receive ZF aims at nulling the interference between sub-arrays by using ZF beamforming at the transmitter and receiver.
Due to zero interference, beamforming and power allocation can be separated.
Hence, the well-known water-filling method is employed to allocate per-stream powers.
A detailed description of the transmit-receive ZF algorithm is given next. 

In this algorithm, the overall combiner ${\bf M}$ is first initialized with $N_s$ left singular vectors of the SVD of channel ${\bf H}$.
Then ZF precoder is calculated seperately for each subarray considering the effective channel ${\bf M}^H {\bf H}_i$ as
\begin{equation}
{\bf V}_i = {\bf H}_i^{H} {\bf M} ({\bf M}^{H} {\bf H}_i {\bf H}_i^{H} {\bf M})^{-1}
\label{ZFprecoder}
\end{equation}
where ${\bf V}_i \in {\mathbb C}^{n \times N_s}$ is the overall precoder corresponding to the $i$th subarray, and ${\bf H}_i \in {\mathbb C}^{M \times n}$ is the channel matrix between the $i$th subarray and the user.
Since sub-array-based processing is used, the $i$th column of ${\bf V}_i$ is picked and normalized as the precoder of the $i$th subarray.
The normalized per-sub-array precoders are stacked into ${\bf {\bar V}} \in {\mathbb C}^{N \times N_s}$, where the entries of other sub-arrays than $i$ are set to zeros in the $i$th column.
The next step is to calculate ZF combiner
with effective channel ${\bf H}\bar{\bf V}$ as
\begin{equation}
{\bf M}^{H} = (\bar{\bf V}^{H} {\bf H}^{ H} {\bf H}\bar{\bf V})^{-1} \bar{\bf V}^{H} {\bf H}^{H}
\label{ZFcombiner}
\end{equation}
and normalize all columns of ${\bf M}$, one by one.
Transmit-receive ZF process is repeated until a desired level of convergence is achieved in terms of rate.

To maximize the rate, waterfilling (WF) algorithm is employed to allocate power $P_i$ to the $i$th stream as
\begin{equation}
P_i = (\mu - \frac{N_0}{\epsilon_i})^{+}
\label{waterfilling}
\end{equation}
where $\mu$ is the water level constant, and the channel gain $\epsilon_i$ for the $i$th stream can be calculated as
\begin{equation}
\epsilon_i = |{\bf w}_{i}^{H} {\bf H} {\bf {\bar v}}_{i}|^2
\label{eigenvalues}
\end{equation}
where ${\bf {\bar v}}_{i}$ and ${\bf w}_{i}$ are the $i$th column of the normalized overall precoder and the digital combiner, respectively.
The total transmit power is limited by $\sum_{i=1}^{N_s} P_i = P$. The overall hybrid precoding matrix is formed as ${\bf V} = \sqrt{\bf P} {\bf {\bar V}}$ where  ${\bf P} = {\rm diag}(P_{1}, P_{2}, \ldots, P_{N_s})$.
The overall transmit-receive ZF method is summarized in Algorithm \ref{transmitreceiveZFalgorithm}.

\section{HBF Algorithms for MU-MISO}
\label{HBFAlgorithmsforMUMISO}

In this section, three partially connected HBF algorithms are developed for MU-MISO systems.
The algorithms are designed for both full- and sub-array-based processing strategies.
In the following, each algorithm is introduced in detail.

\subsection{Full Array-Based Hybrid WMMSE}

This algorithm solves the original sum rate maximization problem in by using alternating optimization on the equivalent problem of weighted MSE minimization so that the objective value converges.
In this algorithm, first, (\ref{WMSEminproblemMUMISO}) is solved with respect to the digital precoder $\{{\bf \bar d}_k\}$ while the analog beamformer is fixed.
Then, the analog beamformer ${\bf A}$ is optimized while keeping the digital precoder fixed.
In the following, the MU-MISO hybrid WMMSE algorithm is described in detail.

Problem (\ref{WMSEminproblemMUMISO}) is convex with respect to the digital precoder $\{{\bf \bar d}_k\}$. The Lagrangian expression of (\ref{WMSEminproblemMat}) is given by
\begin{equation}
\begin{split}
\mathcal{L}
& = \sum_{k=1}^{N_u} w_k  - \sum_{k=1}^{N_u} w_k  m_k {\bf h}_k^H {\bf A} {\bf \bar d}_k  - \sum_{k=1}^{N_u} w_k  {\bf \bar d}_k^H {\bf A}^H {\bf h}_k m_k^{\ast}\\
& \hspace{7mm} +  \sum_{k=1}^{N_u} w_k  m_k {\bf h}_k^H {\bf A} \sum_{j=1}^{N_s} {\bf \bar d}_j {\bf \bar d}_j^H {\bf A}^H {\bf h}_k m_k^{\ast}\\
& \hspace{7mm}  + \sum_{k=1}^{N_u} w_k m_k m_k^{\ast} N_0 + \alpha (\sum_{j=1}^{N_s} {\rm tr}\,({\bf A} {\bf \bar d}_j {\bf \bar d}_j^{H} {\bf A}^{H}) -  P)
\end{split}
\label{LagrangianFullMUMISO}
\end{equation}
The resulting expression from the first order optimality condition is
\begin{equation}
{\bf \bar d}_k = ({\bf A}^H \sum_{j=1}^{K} {\bf h}_j m_j^{\ast} w_j m_j {\bf h}_j^H {\bf A} + \alpha {\bf A}^H {\bf A})^{-1} {\bf A}^H {\bf h}_k m_k^{\ast} w_k
\label{ckFullMUMISO}
\end{equation}
where $\alpha \geq 0$ is chosen such that the transmit power constraint is satisfied and can be found using the bisection method.
In the next step, we fix the digital precoder and solve (\ref{WMSEminproblemMUMISO}) with respect to the analog beamformers for different subarrays.
The received signal of the $k$th user can be rewritten as
\begin{equation}
\begin{split}
y_k
&= \sum\limits_{\substack{j=1}}^{N_s} {\bf h}_{jk}^{H} {\bf a}_j d_{jk} s_{k} + \sum\limits_{\substack{j=1}}^{N_s} {\bf h}_{jk}^{H} {\bf a}_j \sum\limits_{\substack{i=1 \\ i \neq k}}^{N_u} d_{ji} s_{i} + z_k
\end{split}
\label{kthreceivedsignalRE}
\end{equation}
and the error term corresponding to the $k$th user as
\begin{equation}
\begin{split}
e_k
& ={\mathbb E}\left[ \left( s_k - m_k y_k \right) \left( s_k - m_k y_k \right)^H \right]\\
&= 1 - m_k \sum_{j=1}^{N_s} {\bf h}_{jk}^H {\bf a}_j d_{jk} - \sum_{j=1}^{N_s} d_{jk}^{\ast} {\bf a}_j^H {\bf h}_{jk} m_k^{\ast}\\
&\hspace{7mm} +  m_k \sum_{j=1}^{N_a} {\bf h}_{jk}^H {\bf a}_j {\bf d}_j \sum_{l=1}^{N_s} {\bf d}_l^{H} {\bf a}_l^H {\bf h}_{lk} m_k^{\ast} + N_0 m_k m_k^{\ast}
\end{split}
\label{ekRE}
\end{equation}
where ${\bf h}_{ik}$ is the channel vector between $i$th subarray and user $k$.
Consequently, the Lagrangian expression corresponding to (\ref{WMSEminproblemVec}) can be expressed as
\begin{equation}
\begin{split}
\mathcal{L}
& = \sum_{k=1}^{N_u} w_k  - \sum_{k=1}^{N_u} w_k m_k \sum_{j=1}^{N_s} {\bf h}_{jk}^H {\bf a}_j d_{jk}\\
& \hspace{7mm} - \sum_{k=1}^{N_u} w_k \sum_{j=1}^{N_s} d_{jk}^{\ast} {\bf a}_j^H {\bf h}_{jk} m_k^{\ast} + \sum_{k=1}^{N_u} w_k N_0 m_k m_k^{\ast}\\
& \hspace{7mm} +  \sum_{k=1}^{N_u} w_k  m_k \sum_{j=1}^{N_a} {\bf h}_{jk}^H {\bf a}_j {\bf d}_j \sum_{l=1}^{N_s} {\bf d}_l^{H} {\bf a}_l^H {\bf h}_{lk} m_k^{\ast}\\
& \hspace{7mm} + \alpha (\sum_{j=1}^{N_a} {\rm tr}\,({\bf a}_j {\bf d}_j {\bf d}_j^{H} {\bf a}_j^{H}) -  P).
\end{split}
\label{LagrangianFullMUMISO}
\end{equation}

\begin{figure*}
\begin{equation}
\begin{split}
{\bf a}_i
& = (\sum_{k=1}^{N_u} {\bf h}_{ik} m_k^{\ast} w_k m_k {\bf h}_{ik}^H + \alpha {\bf I}_n)^{-1} \sum_{k=1}^{N_u} {\bf h}_{ik} m_k^{\ast} w_k \bigg( d_{ik}^{\ast} - m_k \sum \limits_{\substack{j=1, j\neq i}}^{N_s} {\bf h}_{jk}^H {\bf a}_j {\bf d}_j {\bf d}_i^H \bigg) ({\bf d}_i {\bf d}_i^H)^{-1}
\end{split}
\label{aiFullMUMISO}
\end{equation}
\end{figure*}

The first order optimality condition with respect to ${\bf a}_i$ yields (\ref{aiFullMUMISO})
in which $\alpha \geq 0$ is chosen via the bisection method such that the transmit power constraint is satisfied.
This alternating optimization procedure is continued until achieving a desired level of convergence.
The proposed algorithm converges in terms of objective value since solving a convex problem at each step improves the objective value and the resulting MSE is lower bounded \cite{ShiRLH}.
Due to the non-convexity of the original problem, optimality of the solution cannot be guaranteed.
Hence, the solution has to be considered as suboptimal unless otherwise proven.
Algorithm \ref{MUMISOFullarrayHybridWMMSEalgorithm} summarizes the proposed full array-based hybrid WMMSE approach.

\begin{algorithm}[t]
  \caption{Full Array-Based Hybrid WMMSE Algorithm}\label{MUMISOFullarrayHybridWMMSEalgorithm}
  \begin{algorithmic}[1]
      \State Set iteration number $n=0$.
      \State Initialize $\{{\bf \bar d}_k\}^{n}$ and ${\bf A}^{n}$.
      \Repeat
      \State Update $n=n+1$.
      \State Compute $\{m_k^{n}\}$ from (\ref{MMSEUserReceiver}) given $\{{\bf \bar d}_k\}^{n-1}$ and\\ \hspace{5mm} ${\bf A}^{n-1}$.
      \State Compute $\{w_k^{n}\}$ from (\ref{wk}), (\ref{ek}) given $\{{\bf \bar d}_k\}^{n-1}$ and\\ \hspace{5mm} ${\bf A}^{n-1}$.
      \State Compute $ \{{\bf \bar d}_k\}^{n}$ from (\ref{ckFullMUMISO}) given ${\bf A}^{n-1}$.
      \State Compute $\{{\bf a}_{i}\}^{n}$ from (\ref{aiFullMUMISO}) given ${\bf D}^{n}$.
      \Until{desired level of convergence}
  \end{algorithmic}
\end{algorithm}

\subsection{Subarray-Based Hybrid WMMSE}

The sub-array-based WMMSE scheme is a reduced version of Algorithm \ref{MUMISOFullarrayHybridWMMSEalgorithm} due to the digital beamformer being fixed as an identity matrix.
In this case, the multi-user WMMSE problem in (\ref{WMSEminproblemMUMISO}) can be solved for the analog beamformer $\{{\bf a}_j\}$.
The received signal of the $k$th user can be rewritten as
\begin{equation}
\begin{split}
y_k
&= {\bf h}_{kk}^{H} {\bf a}_k s_{k} + \sum\limits_{\substack{i=1, i \neq k}}^{N_s} {\bf h}_{ik}^{H} {\bf a}_i s_{i} + z_k
\end{split}
\label{kthreceivedsignalRE}
\end{equation}
and the error term corresponding to the $k$th user as
\begin{equation}
\begin{split}
e_k
& ={\mathbb E}\left[ \left( s_k - m_k y_k \right) \left( s_k - m_k y_k \right)^H \right]\\
&= 1 - m_k {\bf h}_{kk}^H {\bf a}_k - {\bf a}_k^H {\bf h}_{kk} m_k^{\ast}\\
&\hspace{7mm} +  \sum\limits_{\substack{i=1}}^{N_s} m_k {\bf h}_{ik}^{H} {\bf a}_i {\bf a}_i^{H} {\bf h}_{ik} m_k^{\ast} + N_0 m_k m_k^{\ast}
\end{split}
\label{ekRE}
\end{equation}
where ${\bf h}_{ik}$ is the channel vector between $i$th subarray and user $k$.
Using the stream specific MSE expressions, the corresponding Lagrangian expression is given by
\begin{equation}
\begin{split}
\mathcal{L} =
& \sum_{k=1}^{N_s} w_k  - \sum_{k=1}^{N_s} m_k w_k {\bf h}_{kk}^H {\bf a}_k - \sum_{k=1}^{N_s} w_k {\bf a}_k^H {\bf h}_{kk} m_k^{\ast}\\
& \hspace{7mm} +  \sum_{k=1}^{N_s} w_k m_k \sum\limits_{\substack{i=1}}^{N_s} {\bf h}_{ik}^{H} {\bf a}_i {\bf a}_i^{H} {\bf h}_{ik} m_k^{\ast}\\
& \hspace{7mm} + N_0 \sum_{k=1}^{N_s} w_k m_k m_k^{\ast} + \alpha (\sum_{i=1}^{N_s} {\rm tr}\,({\bf a}_i {\bf a}_i^{H}) -  P).
\end{split}
\label{LagrangianSub}
\end{equation}
Solving this for the $k$th user receiver $m_k$ result in
\begin{equation}
m_k = {\bf a}_k^H {\bf h}_{kk} (\sum_{i=1}^{N_a} {\bf h}_{ik}^H {\bf a}_i {\bf a}_i^H {\bf h}_{ik} + N_0)^{-1}.
\label{mksSub}
\end{equation}
After applying the MMSE receiver, the error term at the $k$th user is given by
\begin{equation}
\begin{split}
e_k
& = 1 - {\bf a}_k^H {\bf h}_{kk} (\sum_{i=1}^{N_s} {\bf h}_{ik}^H {\bf a}_i {\bf a}_i^H {\bf h}_{ik} + N_0)^{-1} {\bf h}_{kk}^H {\bf a}_k
\end{split}
\label{ekafterSub}
\end{equation}

\begin{algorithm}[t]
  \caption{Subarray-Based Hybrid WMMSE Algorithm}\label{MUMISOSubarrayHybridWMMSEalgorithm}
  \begin{algorithmic}[1]
      \State Set iteration number $n=0$.
      \State Initialize ${\bf A}^{n}$.
      \Repeat
      \State Update $n=n+1$.
      \State Compute $\{m_k^{n}\}$ from (\ref{mksSub}) given ${\bf A}^{n-1}$.
      \State Compute $\{w_k^{n}\}$ from (\ref{wk}), (\ref{ek}) given ${\bf A}^{n-1}$.
      \State Compute $\{{\bf a}_{i}\}^{n}$ from (\ref{aisubMUMISO}).
      \Until{desired level of convergence}
  \end{algorithmic}
\end{algorithm}

The first order optimality condition of $\mathcal{L}$ with respect to each ${\bf a}_i$ yields
\begin{equation}
{\bf a}_i = (\sum_{k=1}^{K} {\bf h}_{ik} m_k^{\ast} w_k m_k {\bf h}_{ik}^H + \alpha {\bf I}_n)^{-1} {\bf h}_{ii} m_i^{\ast} w_i
\label{aisubMUMISO}
\end{equation}
where $\alpha \geq 0$ is chosen via the bisection method while satisfying the transmit power constraint.
This alternating optimization procedure is repeated until a desired level of convergence is obtained.
The developed subarray-based hybrid WMMSE approach is summarized in Algorithm \ref{MUMISOSubarrayHybridWMMSEalgorithm}.
The convergence and suboptimality properties of Algorithm \ref{MUMISOFullarrayHybridWMMSEalgorithm} apply also for Algorithm \ref{MUMISOSubarrayHybridWMMSEalgorithm}.


\subsection{Subarray-Based ZF}
In the subarray-based ZF method, each subarray uses ZF beamformer to its corresponding user while nulling the interference to the other users.
To cancel the interference successfully in this method the number of antennas of each subarray is required to be more than or equal to the number of users.
Stacking the channel vectors, the normalized subarray-based ZF beamformer can be formed as
\begin{equation}
{\bf {\bar V}}
=\left(
\begin{array}{cccc}
\frac{[{\bf V}_1]_1}{\left\Vert[{\bf V}_1]_1\right\Vert} & {\bf 0} &  \ldots & {\bf 0}\\
{\bf 0} & \frac{[{\bf V}_2]_2}{\left\Vert[{\bf V}_2]_2\right\Vert} & \ldots & {\bf 0}\\
\vdots & \vdots & \ddots & \vdots\\
{\bf 0} & {\bf 0} & \ldots & \frac{[{\bf V}_K]_K}{\left\Vert[{\bf V}_K]_K\right\Vert}
\end{array} \right)
\label{subarrayZF}
\end{equation}
where ${\bf V}_k = {\bf H}_k^H ({\bf H}_k {\bf H}_k^H)^{-1}$. The matrix ${\bf H}_k \in {\mathbb C}^{n \times K}$ is the channel between $k$th subarray and all $K$ users, and $[.]_k$ denotes the $k$th column of its matrix argument.
Then the overall beamformer is formed as ${\bf V} = \sqrt{\bf P} {\bf {\bar V}}$ in which WF power allocation is used to form ${\bf P}$. The complexity of this algorithm is considerably less than previous WMMSE based hybrid algorithms.

\section{Simulation Results}
\label{SimulationResults}

In this section, the performance of the proposed HBF algorithms is evaluated via a set of numerical examples in SU-MIMO and MU-MISO scenarios.
The propagation channel between the BS and each user is constructed using two different channel models.
The first one is a geometric uniform linear array (ULA) antenna configuration, also known as Saleh-Valenzuela model \cite{ARAPH}.
The second one is a geometry-based stochastic channel model, called the New York University Simulator (NYUSIM) \cite{RXMMMZ, SR}, which is based on channel measurements at mm-wave frequencies.
In the SU-MIMO scenario, the HBF algorithms are compared to optimal fully digital and analog beamforming solutions, i.e., multi- and single-stream SVD-based approaches, respectively.
In the MU-MISO case, the proposed HBF schemes are compared to fully digital WMMSE and ZF beamforming algorithms.
In the simulations, we use a relatively large number of transmit antennas to reflect the nature of a massive MIMO type of setting and the communication channels are considered as frequency flat.
System performance is measured in terms of achievable rate (i.e., spectral efficiency).
The rate performance is averaged over 100 channel realizations.

\subsection{Performance Evaluation in Geometric ULA Channel}

This subsection provides a performance study of the HBF methods in a geometric channel model with L propagation paths and ULA antenna configuration.
A detailed description of the channel model is given first.
The convergence behavior of the hybrid WMMSE algorithms is studied next.
Then, all the proposed HBF methods are compared to digital and analog beamforming in SU-MIMO and MU-MISO scenarios with various parameter settings.

\begin{figure}
	\begin{subfigure}{0.5\textwidth}
		\centering
		\includegraphics[width=.84\linewidth]{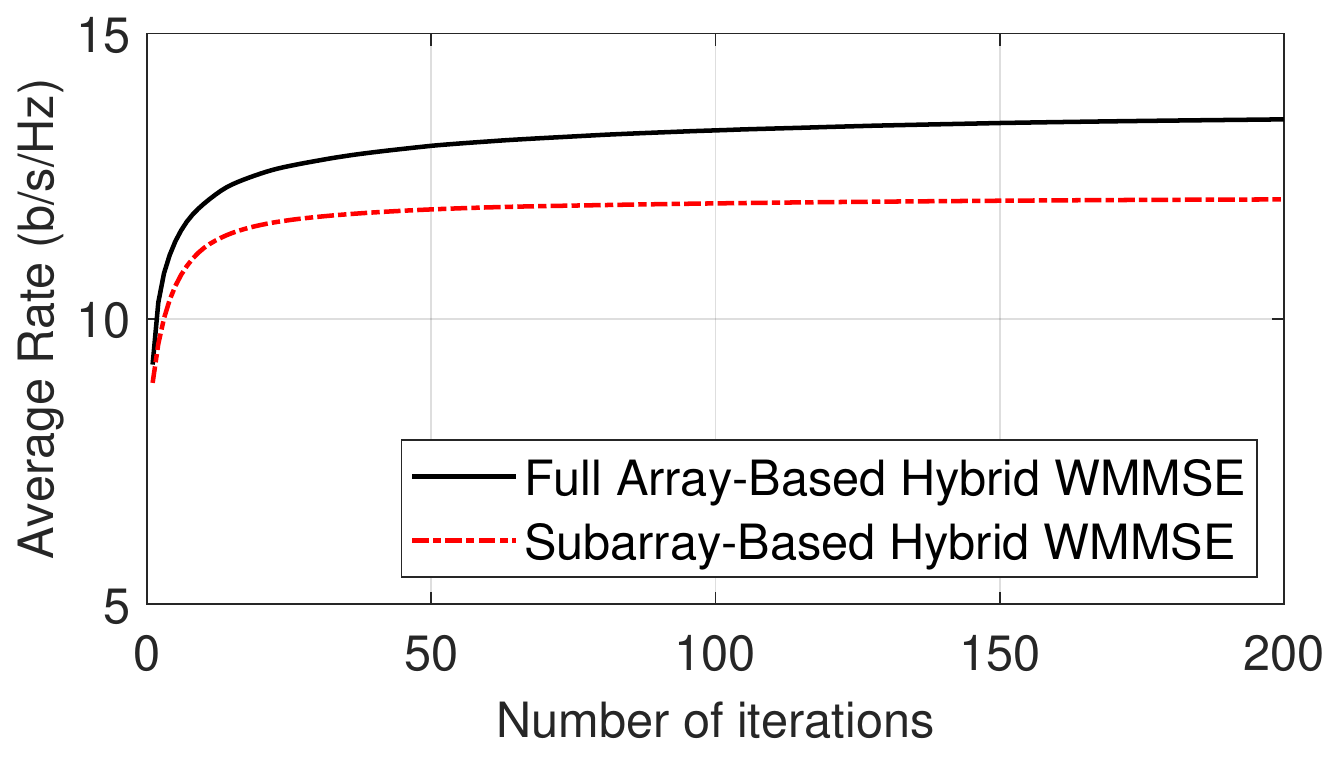}
		\caption{SU-MIMO  with $N_t = 64$, $N_r = 4$, $N_s = 4$.}
		\label{ConvergenceSU00dB5e2}
	\end{subfigure}\\
	\begin{subfigure}{0.5\textwidth}
		\centering
		\includegraphics[width=.84\linewidth]{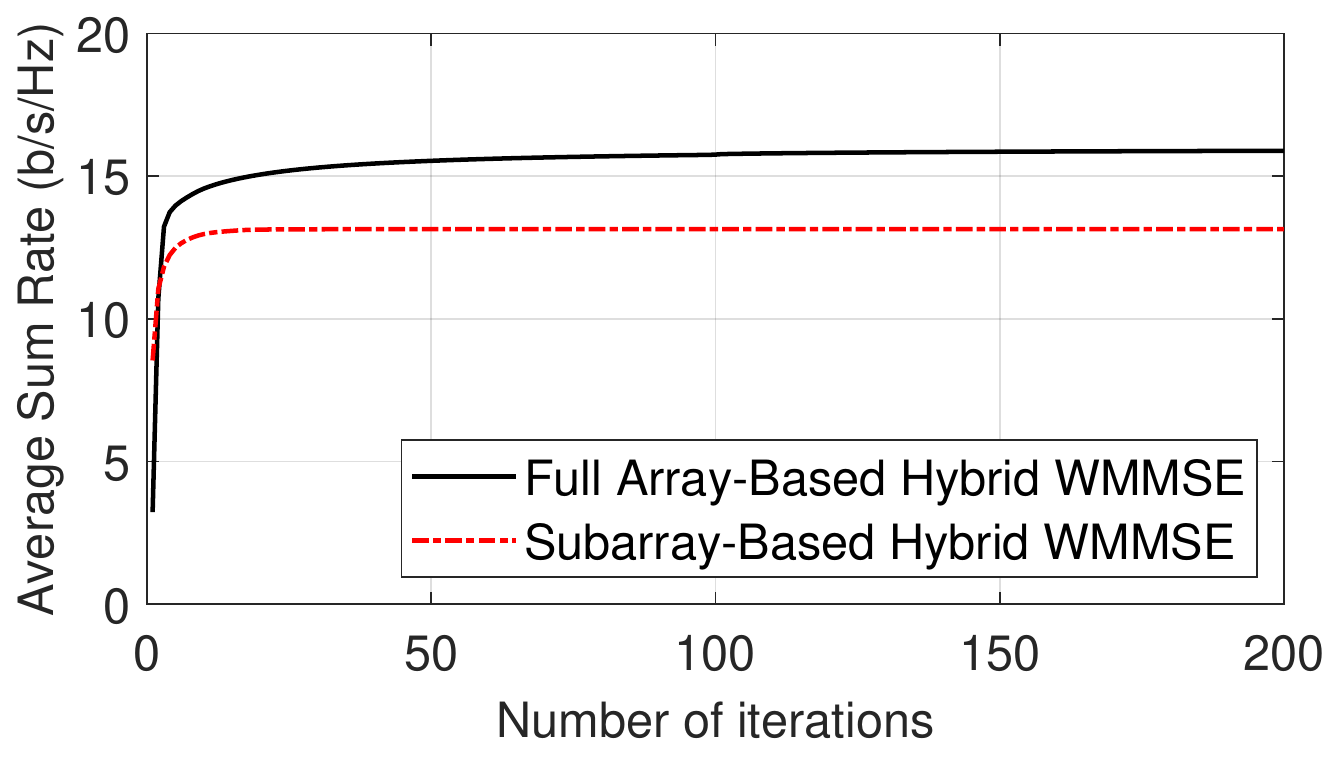}
		\caption{MU-MISO  with $N_t = 64$, $N_u = 4$.}
		\label{ConvergenceSU20dB5e3}
	\end{subfigure}
	\caption{ Convergence behavior of WMMSE algorithms at $5$ dB SNR.}
	\label{Convergence}
\end{figure}


The MIMO channel matrix corresponding to the geometric ULA is expressed as
\begin{equation}
{\bf H} = \sqrt{\frac{N_t N_r}{L}} \sum_{l=1}^{L} \alpha_l {\bf g}_r (\phi_r^l) {\bf g}_t (\phi_t^l)^H
\label{ULAchannel}
\end{equation}
where $L$ is the number of paths between the BS and the user, $\alpha_l \sim \mathcal{CN}(0,1)$ is the path gain for the $l$th path, $\phi_r^l \in [0, 2\pi)$ and $\phi_t^l \in [0, 2\pi)$ are the angles of arrival and departure for the $l$th path, respectively.
Moreover, ${\bf g}_r (\phi_r^l)$ and ${\bf g}_t (\phi_t^l)$ are the receive and transmit antenna array response vectors, respectively.
These array response vectors are functions of angles of 
The ULA is considered to have $L=5$ paths with uniform distribution for both angles of departure and arrival, and the antenna spacing equal to half a wavelength.
Moreover, the channel is assumed to be frequency flat with normalized path loss.
The array response vector of an $N_t$-element ULA can be written as:
\begin{equation}
{\bf g} (\phi) = \frac{1}{\sqrt{N_t}} \left( 1,\: e^{jkd\,sin(\phi)},\: ...,\: e^{j(N_t -1)kd\,sin(\phi)}  \right)^T
\label{arrayresponsevec}
\end{equation}
where $k=2\pi f$ and $d$ is the antenna inter-element spacing. In this paper, the carrier frequency is considered to be $f=28\: \mathrm{GHz}$ to reflect the mm-wave communications specification.


Fig. \ref{Convergence} shows the convergence behavior of the hybrid WMMSE algorithms in ULA channel model.
It can be seen that the convergence rate of the algorithms in both SU and MU cases is relatively fast at the first tens of iterations.
The subarray based algorithms approximately converge after only tens of iterations, however, the convergence speed of the full-array based algorithms become slow closer to the convergence value.

\subsubsection{SU-MIMO}

\begin{figure}
	\centering
	\includegraphics[width=.94 \linewidth]{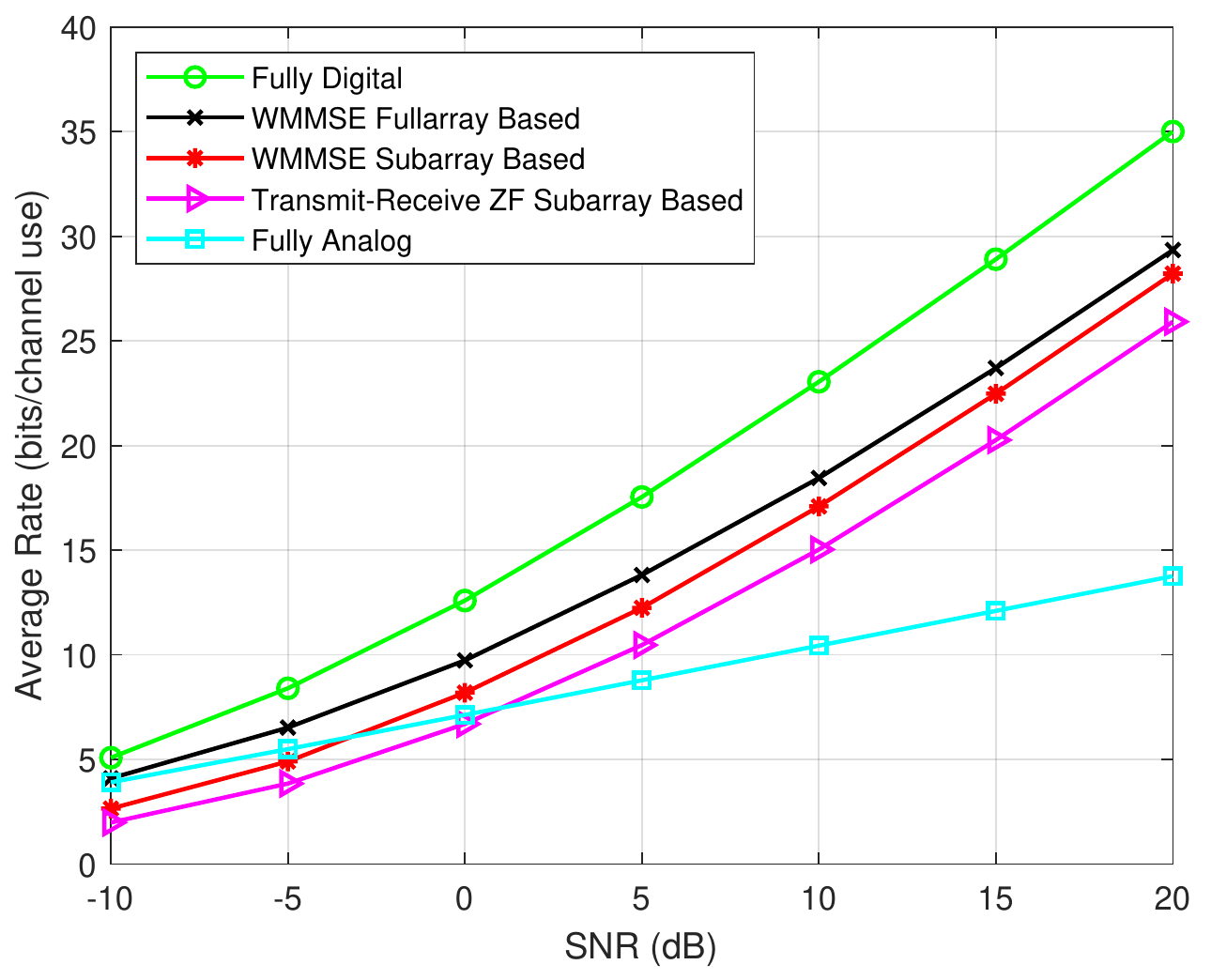}
	\caption{Rate vs. SNR in SU-MIMO system with $N_t = 64$, $N_r = 4$, $N_s = 4$.}
	\label{RatevsSNR}
\end{figure}

\begin{figure}
	\centering
	\includegraphics[width=.94 \linewidth]{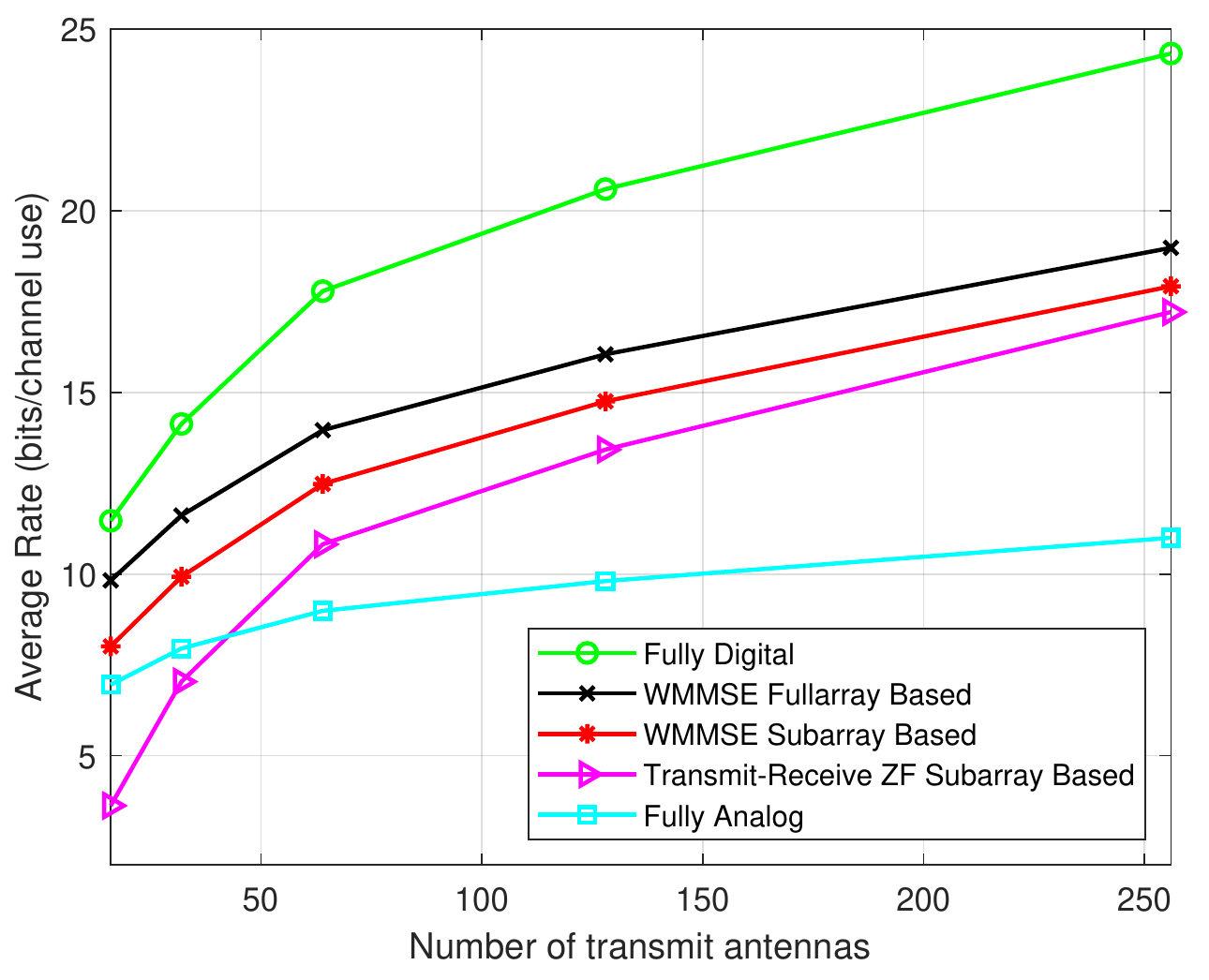}
	\caption{Rate vs. number of transmit antennas in SU-MIMO system with $N_r = 4$, $N_s = 4$ at ${\rm SNR}=5$ dB.}
	\label{RatevsNt}
\end{figure}

Fig. \ref{RatevsSNR} illustrates the average rate versus SNR in geometric ULA channel model.
The numbers of transmit antennas, receive antennas, and data streams are considered to be $64$, $4$, and $4$, respectively.
As it can be seen, full-array based hybrid WMMSE algorithm achieves the best rate performance among all other hybrid algorithms all over the SNR range.
At lower SNRs, its performance is close to the optimal fully digital solution while a moderate gap appears in between them at higher SNRs.
This gap is mainly due to the partially connected hybrid architecture with lower analog RF chains and consequently lower array gains.
Subarray based HBF algorithm's rate performance is slightly lower than its full-array based counterpart.
TxRxZF algorithm performs slightly lower than subarray based hybrid scheme all over SNR regime, however, its performance is lower than full-array based hybrid scheme with a moderate gap which is mainly due to subarray based design.
Fully analog (one stream) scheme has the lowest performance in all over the SNR range except in very low SNRs in which TxRxZF, and subarray based hybrid algorithms drop below fully analog scheme.


Fig. \ref{RatevsNt} presents the average rate versus number of transmit antennas.
The numbers of receive antennas and data streams are considered to be 4 and the performance is evaluated at SNR=$5$ dB.
One can see that the rate performance of the full-array based hybrid WMMSE design is superior to all other hybrid algorithms, but inferior to fully digital solution all over the number of transmit antenna range.
The second best among hybrid algorithms is subarray based hybrid WMMSE scheme and the last is TxRxZF design, each with a small gap from the curve with better performance.

\subsubsection{MU-MISO}

\begin{figure}
\centering
\includegraphics[width=.94 \linewidth]{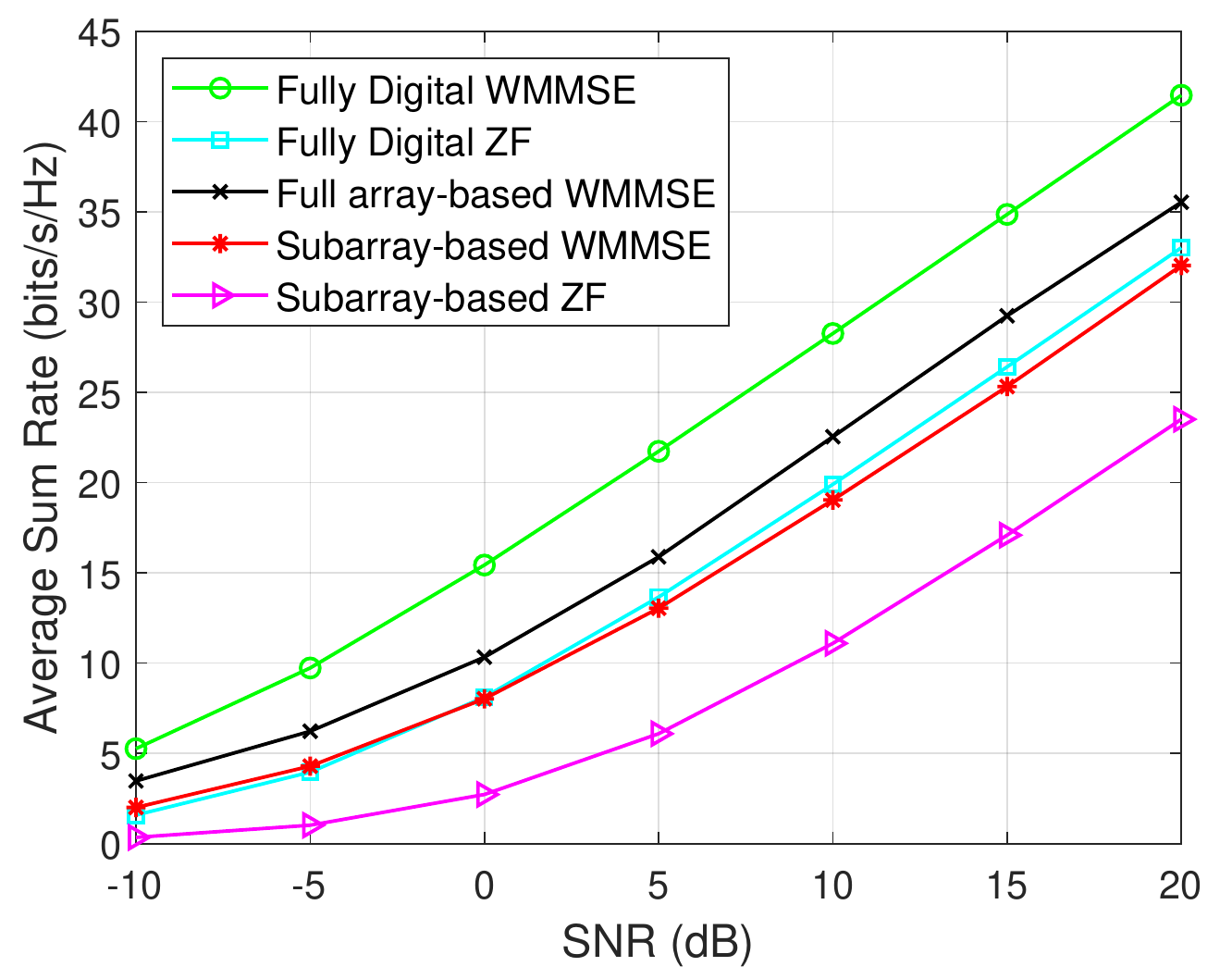}
\caption{Sum rate vs. SNR in MU-MISO system with $N_t = 64$, $N_u = 4$.}
\label{SumratevsSNR}
\end{figure}

Fig. \ref{SumratevsSNR} shows the average sum rate versus SNR in for MU-MISO algorithms in geometric ULA channel model.
The numbers of transmit antennas and users are considered to be $64$ and $4$, respectively.
One can see that the performance of full-array based hybrid WMMSE algorithm is the best among all other hybrid algorithms all over the SNR range. There is a gap between the performance of full array based hybrid scheme and fully digital WMMSE solution which is due to the partially connected design.
Fully digital ZF solution has a lower performance compared to fully digital MMSE which is due to the fundamental difference between WMMSE and ZF schemes where WMMSE design can exploit the interference to achieve a better rate performance than ZF.
Subarray based hybrid WMMSE and fully digital ZF algorithms perform approximately the same all over the SNR range.
Subarray based ZF performance is inferior to all other algorithms which is due to both partial connectivity and ZF nature.

\begin{figure}
\centering
\includegraphics[width=.94 \linewidth]{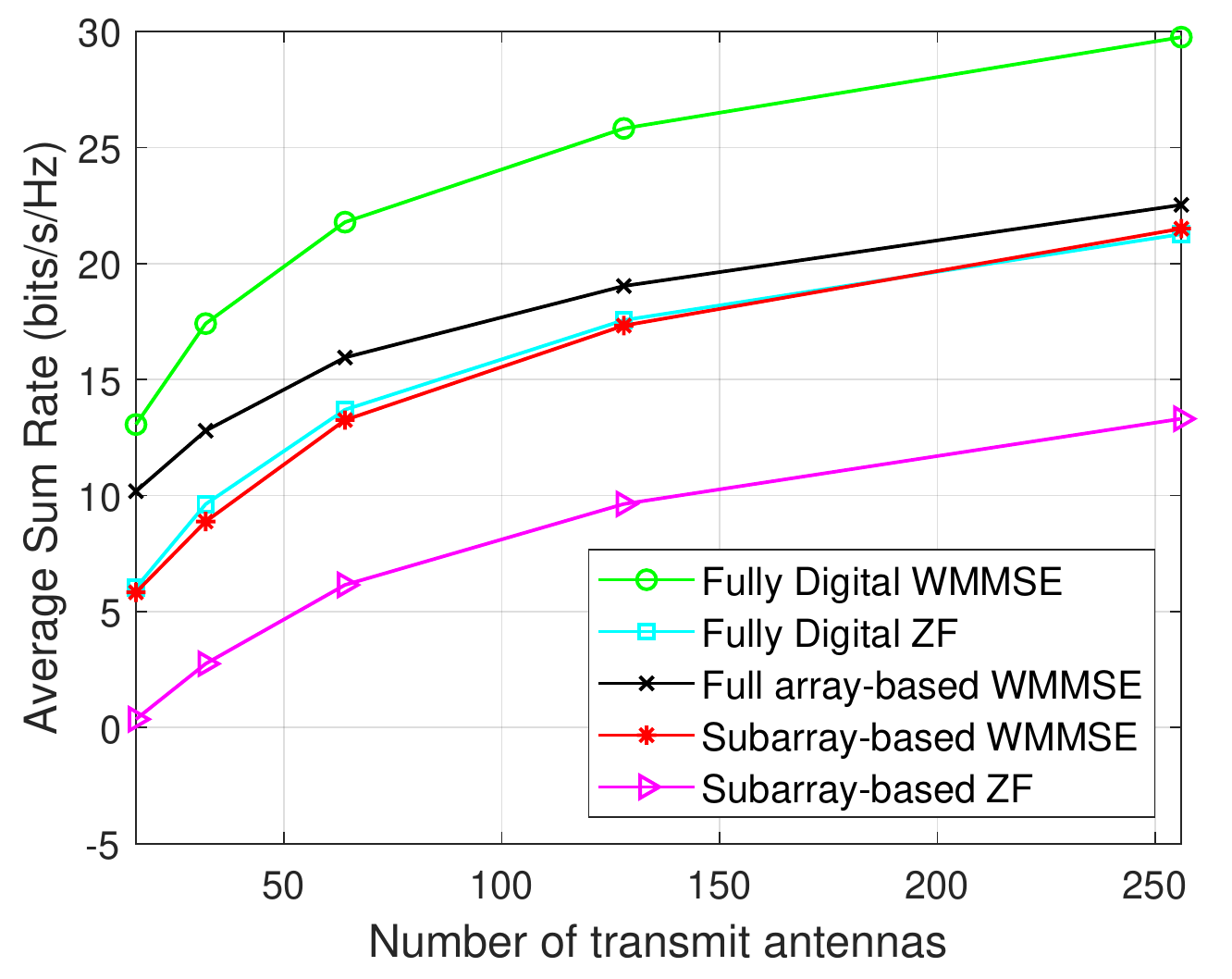}
\caption{Sum rate vs. number of transmit antennas in MU-MISO system with $N_u = 4$, ${\rm SNR}=5$ dB.}
\label{SumratevsNt}
\end{figure}

Fig. \ref{SumratevsNt} illustrates the average sum rate versus number of transmit antennas in MU-MISO system.
The numbers of users is considered to be 4 and the performance is evaluated at $5$ dB SNR.
It can be seen that the full-array based hybrid WMMSE design performs better than all other hybrid algorithms, but inferior to fully digital WMMSE solution all over the number of transmit antenna range.
In higher numbers of transmit antennas, the sum rate increment pace of all the algorithms gets slower and comes to a saturation.
Subarray based hybrid WMMSE and fully digital ZF designs have almost identical performance except all over the range of number of transmit antennas.
Subarray based ZF algorithm is inferior to all hybrid algorithms with a considerable gap.

\subsection{Performance Evaluation in NYUSIM Channel model}

This section evaluates the performance of the HBF schemes in the NYUSIM channel model using a UMa environment with LOS and NLOS propagation conditions.
Various settings of parameters are used for SU-MIMO and MU-MISO scenarios.
Before the performance analysis, the NYUSIM channel model with its main parameters is introduced. 

NYUSIM is a channel simulator developed by New York University which is based on measurements and analysis of the data obtained from various outdoor environments at frequencies from 28 to 73 GHz \cite{RXMMMZ, SR}.
This simulator provides an accurate rendering of both time and space actual channel impulse responses. NYUSIM also provides realistic measured signal levels that can be utilized in realistic physical layer and link layer simulations.
NYUSIM is applicable in carrier frequencies from $500$ MHz to $100$ GHz, and RF bandwidths from 0 (continuous wave) to 800 MHz.

The parameters which are used to evaluate the performance in NYUSIM environment are as follows.
The center frequency of $28$ GHz is used and total transmit power is $10$ W. Urban macrocell (UMa) environment in both outdoor line of sight (LOS) and non-line of sight (NLOS) scenarios are considered.
The transmit and receive antenna array types are considered to be uniform rectangular array (URA) and uniform linear array (ULA).

\subsubsection{SU-MIMO}

\begin{figure}[]
	\begin{subfigure}{0.5\textwidth}
		\centering
		\includegraphics[width=.94 \linewidth]{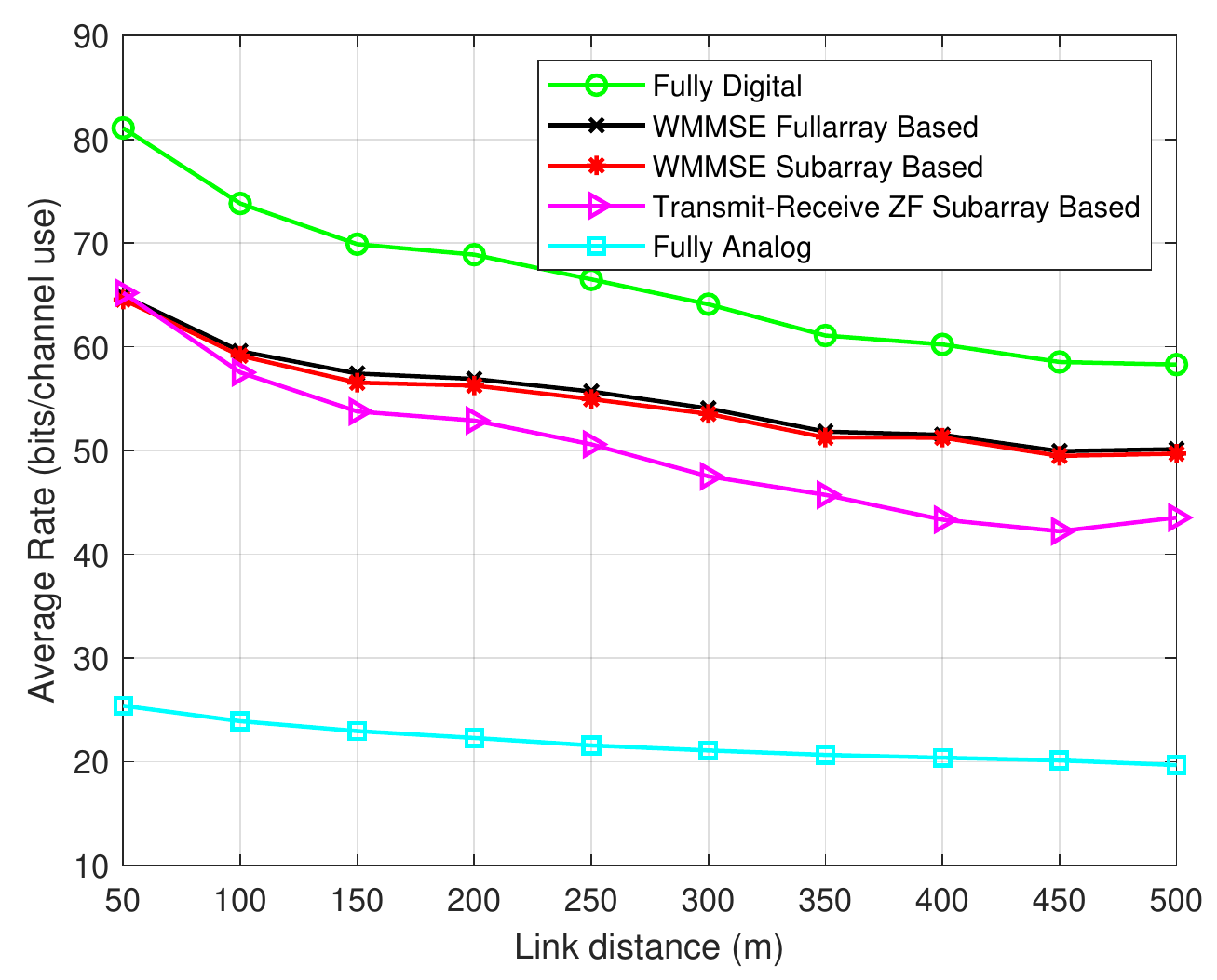}
		\caption {UMa LOS}
		\label{RatevsCellRUMaLOS}
	\end{subfigure}\\
	\begin{subfigure}{0.5\textwidth}
		\centering
		\includegraphics[width=.94 \linewidth]{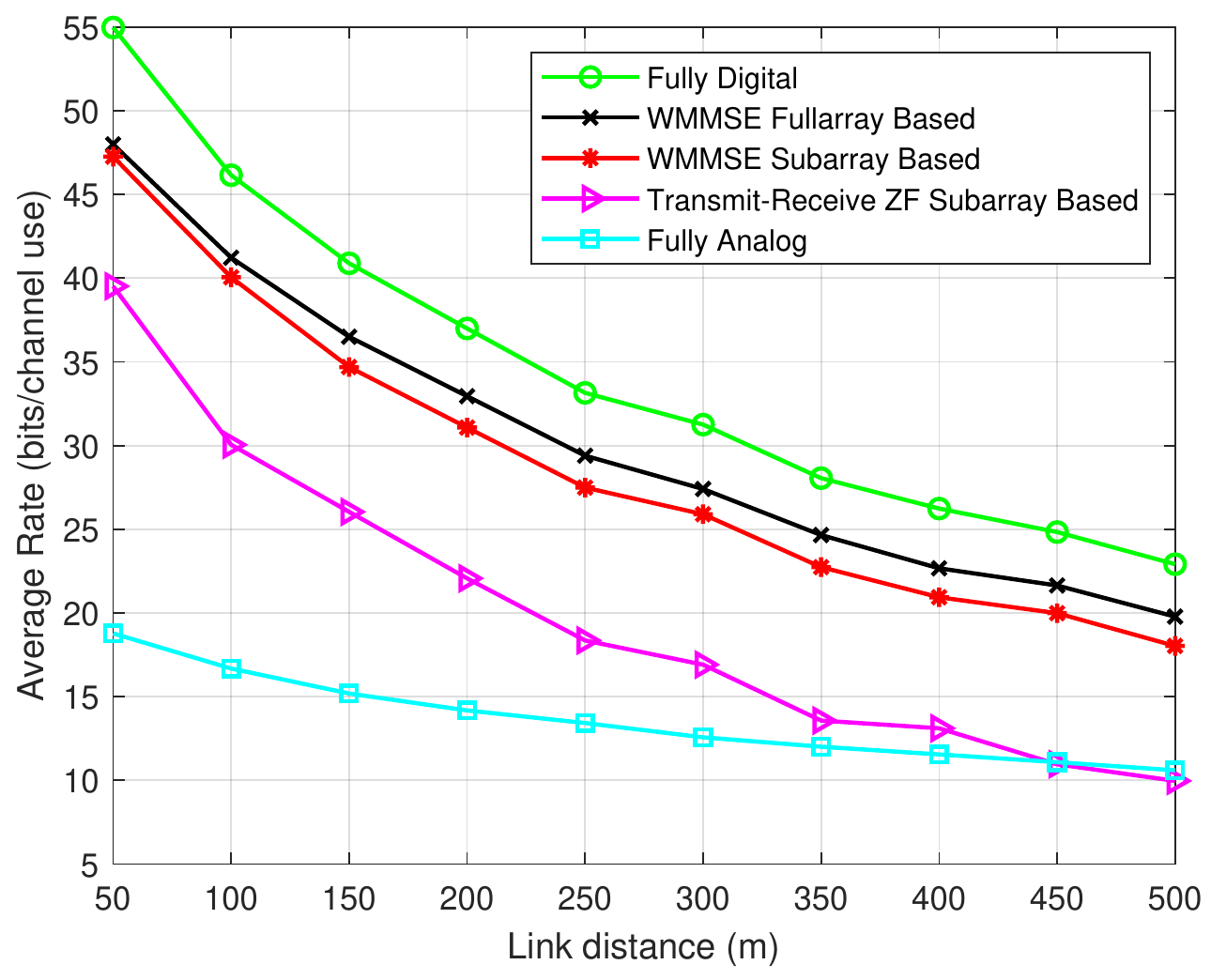}
		\caption {UMa NLOS}
		\label{RatevsCellRUMaNLOS}
	\end{subfigure}
	\caption{Rate vs. link distance in SU-MIMO system with $N_t = 64$, $N_r = 4$, $N_s = 4$.}
	\label{RatevsCellRUMa}
\end{figure}

Fig. \ref{RatevsCellRUMaLOS} and \ref{RatevsCellRUMaNLOS} show the average rate performance against link distance in NYUSIM UMa environment for LOS and NLOS scenarios, respectively.
The numbers of transmit and receive antennas and data streams are set to $64$, $16$, and $4$, respectively and the evaluation is at $28$ GHz carrier frequency.
One can see that the average rate decreases while increasing the link distance  which is due to the path loss.
In both figures, full-array based hybrid WMMSE algorithm achieves the best performance among all hybrid designs.
Subarray based hybrid scheme is close to the full-array counterpart in low cell sizes where in higher radii drops slightly.
The rate decrease of fully analog scheme when the link distance increases is low.
In LOS scenario, the rate performance is considerably higher than the NLOS case which is due to the severe attenuation in mm-wave frequencies. Moreover, the decreasing slope in LOS case is much lower than the NLOS scenario.

\subsubsection{MU-MISO}

\begin{figure}
	\begin{subfigure}{0.5\textwidth}
		\centering
		\includegraphics[width=.94 \linewidth]{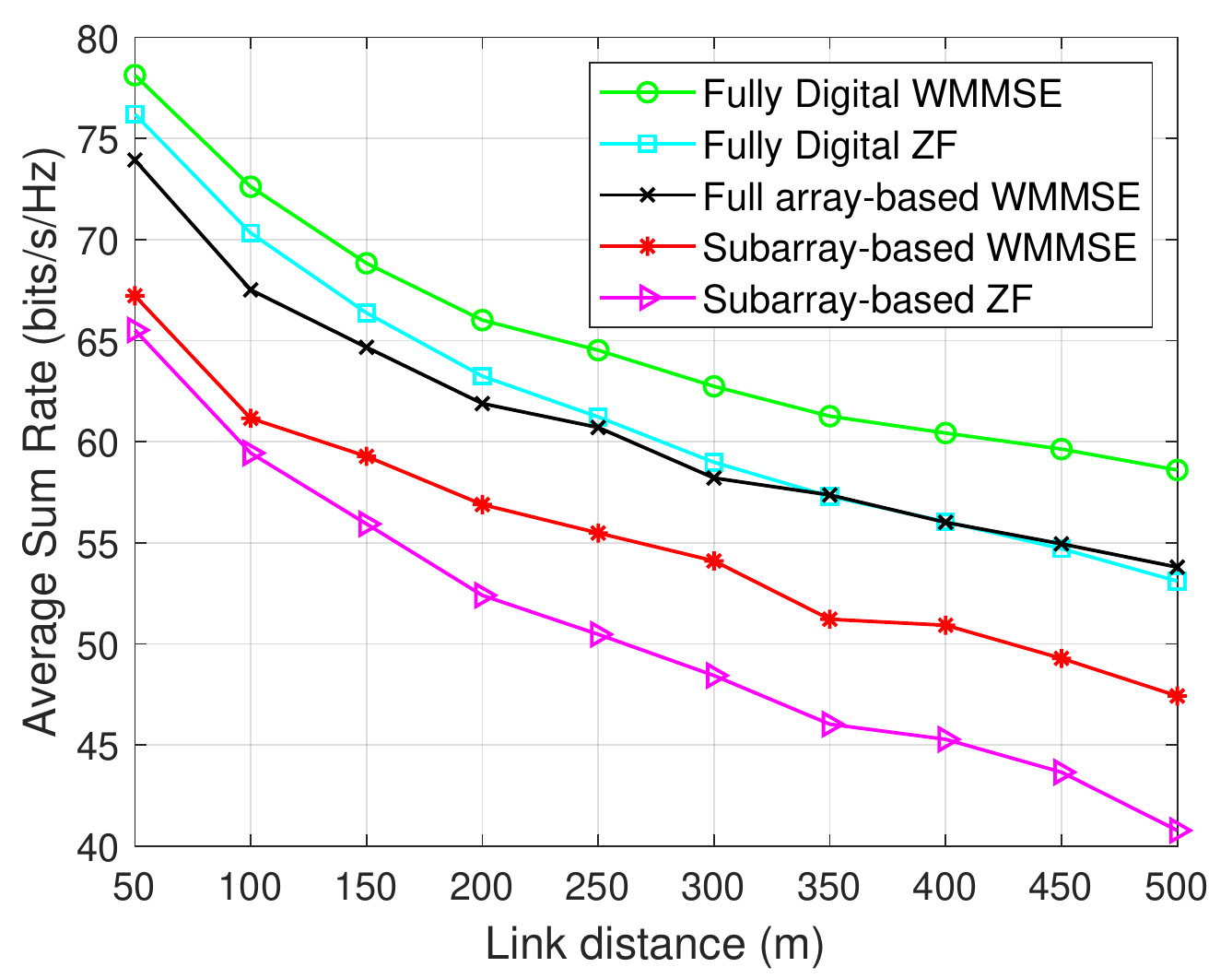}
		\caption {UMa LOS}
		\label{SumratevsCellRUMaLOS}
	\end{subfigure}\\
	\begin{subfigure}{0.5\textwidth}
		\centering
		\includegraphics[width=.94 \linewidth]{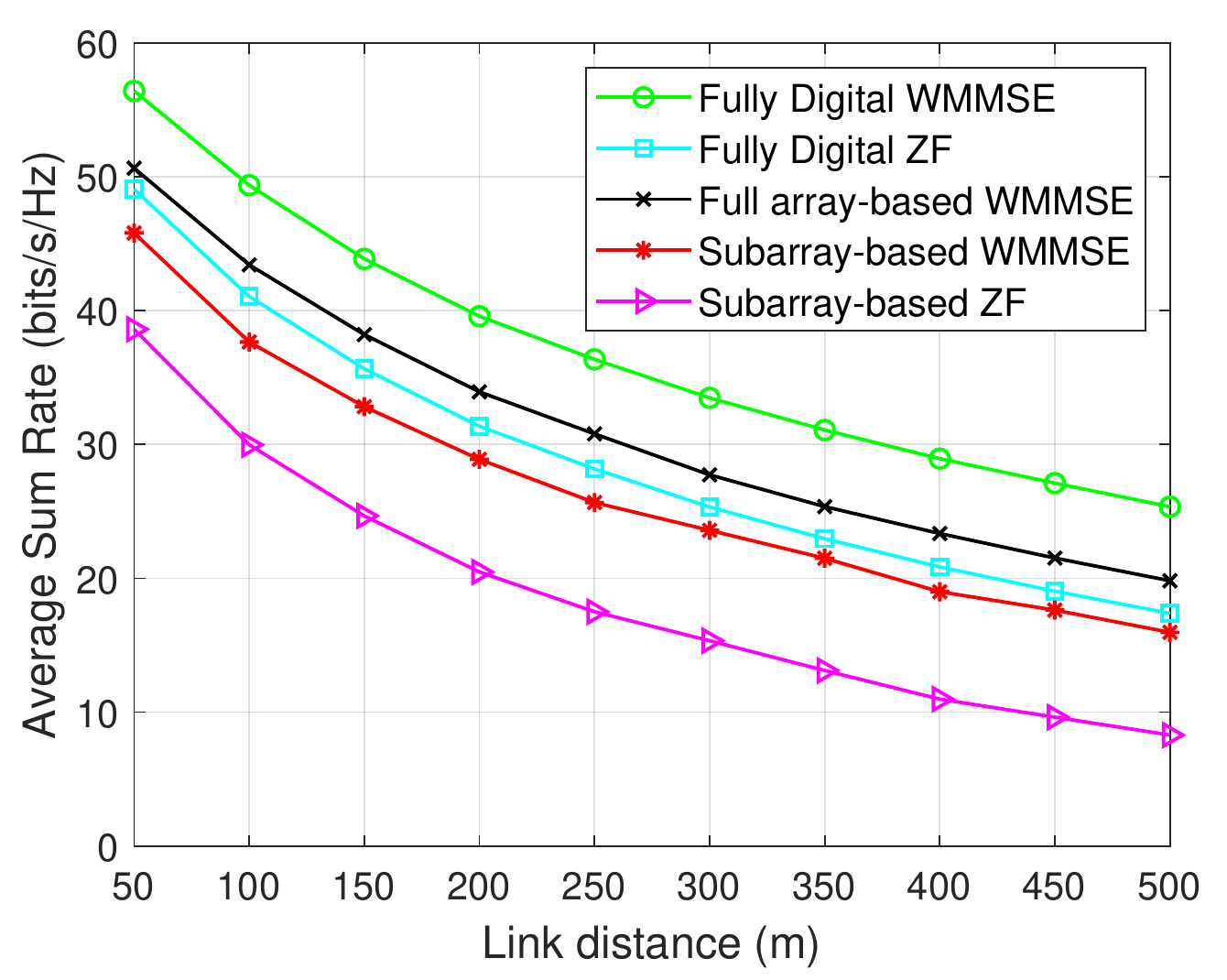}
		\caption {UMa NLOS}
		\label{SumratevsCellRUMaNLOS}
	\end{subfigure}
	\caption{Sum rate vs. link distance in MU-MISO system with $N_t = 64$, $N_u = 4$.}
	\label{SumratevsCellRUMa}
\end{figure}

Fig. \ref{SumratevsCellRUMaLOS} and \ref{SumratevsCellRUMaNLOS} present the average sum rate performance versus link distance in NYUSIM UMa environment for LOS and NLOS scenarios, respectively.
The numbers of transmit antennas and users are set to $64$ and $4$, respectively and the evaluation is at $28$ GHz carrier frequency.
It can be seen that the average sum rate has a decreasing behavior while the link distance increases due to different attenuation.
In both figures, the sum rate performance of full-array based hybrid WMMSE algorithm is the best among all hybrid schemes.
The performance of fully digital ZF solution is slightly better than full-array hybrid algorithm in lower cell sizes of LOS case and slightly lower than it all over all cell sizes of NLOS scenario.
Subarray based ZF algorithm performs inferior to all other designs.
In LOS scenario, the rate performance is considerably higher than the NLOS case which is due to the severe attenuation in mm-wave frequencies.
In NLOS scenario, the decrease in sum rate is considerably higher than the LOS case.

\section{Conclusion}
\label{Conclusion}

In this paper, HBF with partially connected RF architecture was studied for mm-wave massive MIMO systems.
We proposed several optimization-based HBF algorithms for SU-MIMO and MU-MISO settings.
Both full array- and subarray-based processing approaches of partially connected HBF were considered.
We first formulated rate maximization problems for single- and multi-user systems as weighted minimum mean square error (WMMSE) and then derived hybrid beamformers as solutions using alternating optimization.
Moreover, we proposed sub-array-based zero-forcing algorithms with lower complexities.
A simple geometric ULA and a practical NYUSIM channel models were used to compare the rate performance of the hybrid algorithms.
Simulation results showed that partially connected HBF can provide a good balance between hardware complexity and performance compared to fully digital and analog beamforming.
The subarray-based hybrid WMMSE algorithm achieves comparable performance to that of the full array-based WMMSE in a SU-MIMO scenario, while being inferior in a MU-MISO setting.
The hybrid WMMSE results can potentially serve as performance upper bounds for lower complexity algorithms.
An interesting future research subject is to design channel estimation-based HBF algorithms which take into account the frequency selective and sparse nature of mm-wave channels.

\ifCLASSOPTIONcaptionsoff
  \newpage
\fi


%
\balance
\bibliography{JournalBIB}{}
\bibliographystyle{IEEEtran}

%




\end{document}